\documentclass[a4paper,fleqn,usenatbib,useAMS]{mnras}

\usepackage{newtxtext,newtxmath}
\usepackage{comment}

\usepackage[T1]{fontenc}
\usepackage{ae,aecompl}
\usepackage{comment}

\usepackage{graphicx}	
\usepackage{adjustbox}
\usepackage{amsmath}	
\usepackage{amssymb}	
\usepackage{color}

\definecolor{brown}{rgb}{0.6,0.4,0.2}
\definecolor{purple}{rgb}{0.5,0,0.5}

\def\msol{\hbox{\kern 0.20em $M_\odot$}}
\newcommand{\lsol}{\hbox{\kern 0.20em $L_\odot$}}
\newcommand{\g}{\hbox{\kern 0.20em g}}
\newcommand{\gmu}{\hbox{\kern 0.20em g$^{-1}$}}
\newcommand{\kg}{\hbox{\kern 0.20em kg}}
\newcommand{\pc}{\hbox{\kern 0.20em pc}}
\newcommand{\mum}{\hbox{\kern 0.20em $\mu$m}}
\newcommand{\mumd}{\hbox{\kern 0.20em $\mu$m$^{-2}$}}
\newcommand{\cm}{\hbox{\kern 0.20em cm}}
\newcommand{\m}{\hbox{\kern 0.20em m}}
\newcommand{\km}{\hbox{\kern 0.20em km}}
\newcommand{\nm}{\hbox{\kern 0.20em nm}}
\newcommand{\s}{\hbox{\kern 0.20em s}}
\newcommand{\h}{\hbox{\kern 0.20em h}}
\newcommand{\smu}{\hbox{\kern 0.20em s$^{-1}$}}
\newcommand{\smd}{\hbox{\kern 0.20em s$^{-2}$}}
\newcommand{\an}{\hbox{\kern 0.20em an}}
\newcommand{\anmu}{\hbox{\kern 0.20em an$^{-1}$}}
\newcommand{\yr}{\hbox{\kern 0.20em yr}}
\newcommand{\yrmu}{\hbox{\kern 0.20em yr$^{-1}$}}
\newcommand{\Myr}{\hbox{\kern 0.20em Myr}}
\newcommand{\Mymu}{\hbox{\kern 0.20em Myr$^{-1}$}}
\newcommand{\K}{\hbox{\kern 0.20em K}}
\newcommand{\pcmu}{\hbox{\kern 0.20em pc$^{-1}$}}
\newcommand{\pcmd}{\hbox{\kern 0.20em pc$^{-2}$}}
\newcommand{\pcmt}{\hbox{\kern 0.20em pc$^{-3}$}}
\newcommand{\kms}{\hbox{\kern 0.20em km\kern 0.20em s$^{-1}$}}
\newcommand{\kmpd}{\hbox{\kern 0.20em km$^{2}$}}
\newcommand{\kpc}{\hbox{\kern 0.20em kpc}}
\newcommand{\cms}{\hbox{\kern 0.20em cm\kern 0.20em s$^{-1}$}}
\definecolor{purple}{rgb}{0.5,0,0.5}

\newcommand{\sbunit}{erg s$^{-1}$ cm$^{-2}$ sr$^{-1}$}
\newcommand{\subsun}{\mbox{$_{\odot}$}}

\newcommand{\mic}{$\mu$m}

\newcommand{\spitzer}{\textit{Spitzer}}
\newcommand{\herschel}{\textit{Herschel}}
\newcommand{\chandra}{\textit{Chandra}}

\title[A Dust Twin of Cas A]{A Dust Twin of Cas A: \\ Cool Dust and 21$\mu$m Silicate Dust Feature in the Supernova Remnant G54.1+0.3}

\author[J. Rho]{J. Rho$^{1,2}$\thanks{Contact e-mail:{jrho@seti.org}}
H.L. Gomez,$^{3}$
A. Boogert,$^{2}$
M.W.L. Smith,$^{3}$
P.-O Lagage,$^{4}$
D. Dowell,$^{5}$ \and
C.J.R. Clark,$^{3}$
E. Peeters,$^{6,1}$
J. Cami$^{6,1}$
\\
$^{1}$SETI Institute, 189 N. Bernardo Ave, Suite 200, Mountain View, CA 94043\\
$^{2}$SOFIA Science Center, NASA Ames Research Center, MS 232, Moffett Field, CA 94035 \\
$^{3}$School of Physics and Astronomy, Cardiff University, Queens Buildings, The Parade, Cardiff, CF24 3AA, UK\\
$^{4}$Paris-Saclay University, Irfu/AIM, CEA, Universit\'e Paris-Saclay, F-9119 Gif-sur Yvette, France\\
$^{5}$Jet Propulsion Laboratory, California Institute of Technology, 4800 Oak Grove Dr., Pasadena, CA 91104\\
$^{6}$Department of Physics and Astronomy, University of 
Western Ontario, London, ON, N6A 3K7, Canada
}

\date{Accepted}

\pubyear{2018}

\begin{document}
\label{firstpage}
\pagerange{\pageref{firstpage}--\pageref{lastpage}}
\maketitle

\begin{abstract}

We present infrared (IR) and submillimeter observations of the Crab-like
supernova remnant (SNR) G54.1+0.3 including 350$\mu$m (SHARC-II), 870$\mu$m
(LABOCA), 70, 100, 160, 250, 350, 500$\mu$m (\herschel) and 3-40$\mu$m
(\spitzer). We detect dust features at 9, 11 and 21\mic\ and a long
wavelength continuum dust component. The 21\mic\ dust coincides with [Ar
II] ejecta emission, and the feature is remarkably similar to that in Cas
A. The IRAC 8\mic\ image including Ar ejecta is distributed in a shell-like
morphology which is coincident with dust features, suggesting that dust has
formed in the ejecta. We create a cold dust map that shows excess emission
in the northwestern shell. We fit the spectral energy distribution of the
SNR using the continuous distributions of ellipsoidal (CDE) grain model of
pre-solar grain SiO$_2$ that reproduces the 21 and 9\mic\ dust features and
discuss grains of SiC and PAH that may be responsible for the 10-13$\mu$m
dust features. To reproduce the long-wavelength continuum, we explore
models consisting of different grains including Mg$_2$SiO$_4$, MgSiO$_3$,
Al$_2$O$_3$, FeS, carbon, and Fe$_3$O$_4$. We tested a model with a
temperature-dependent silicate absorption coefficient. We detect cold dust
(27-44\,K) in the remnant, making this the fourth such SNR with
freshly-formed dust. The total dust mass in the SNR ranges from
$0.08-0.9\,M_{\odot}$ depending on the grain composition, which is
comparable to predicted masses from theoretical models. Our estimated dust
masses are consistent with the idea that SNe are a significant source of
dust in the early Universe.

\end{abstract}
%




\section{Introduction}

Supernova shocks are known to destroy dust grains in the diffuse
interstellar medium (ISM) and theoretical models of shock-grain
interactions suggests this process is highly efficient, with only a few
percent of the dust mass surviving such an interaction
\citep{mckee87,jones94,bianchi07,bocchio16}. However, it is difficult for
the `textbook' dust-forming scenario, believed to be during the stellar
wind phase of AGB stars, to explain the formation of large amounts of dust
observed in the early Universe
\citep{morgan03,maiolino04,dwek07,michalowski10,watson15,michalowski15,
mattsson15}, unless high star formation rates and highly efficient
theoretical models of dust yields from AGB stars are invoked (e.g.
\citealp{valiante09} and \citealp{ventura12}). Massive star SN events e.g.
Type II, Ib and Ic, could solve this problem since they evolve on much
shorter timescales and eject a large amount of heavy elements into the ISM.
If a typical Type II supernova were able to condense only 10\% of its heavy
elements into dust grains, it could eject around half a solar mass of dust
into the ISM \citep{nozawa03,todini01}. For a star formation rate 100 times
that of the Milky Way this would lead to a dust mass of order $10^8 \rm
M\subsun$ at $z$= 6.3-7.5, for 680-850 million years after the Big Bang, if
SN are not efficient dust destroyers \citep{michalowski15}.

A number of comprehensive studies of nearby SNe have been undertaken in
order to estimate the mass of newly formed dust in the ejecta of
core-collapse SNe (see the review in \citealp{gomez13}). Typical values
derived for very young SNe (within $\sim$ few years after the explosion)
are only $10^{-4}- 10^{-2}M_{\sun}$ of {\it ejecta} dust per SN event (see
for example
\citealp{sugerman06,kotak09,meikle11,andrews11,gall14,gomez14}). In
contrast, the case for significant amounts of dust formed in SN ejecta was
strengthened by mid-IR (\spitzer), FIR (\herschel, AKARI and BLAST) and
submillimeter (submm; SCUBA) observations of much older supernova remnants
(SNRs, $>300$\,years) including Cas A, the Crab Nebula, and SNR
1E0102.2-7219. Using the \spitzer\ mid-IR data alone, a warm dust mass of
$0.02-0.054\,M_{\odot}$ was observed in Cas A \citep{rho08},
3$\times$10$^{-3}$ - $0.014\,M_\odot$ in SNR 1E0102.2-7219
\citep{sandstrom08, rho09}, and $0.012\,M_{\odot}$ in the Crab Nebula
\citep{temim12}.

Adding the longer wavelength FIR-submm observations for
the Cas A remnant revealed a cooler component of dust at $\sim 35\,$K with
mass $0.06-0.5\, M_{\odot}$ \citep{barlow10,sibthorpe10,deLooze17}.
\citet{dunne03} also found a cold dust component radiating at $\sim$20\,K.
Although the emission from these grains are difficult to disentangle from
foreground and background dust structures in the Milky Way \citep{krause04},
\citet{dunne09} showed that a significant mass of the cold dust seen
towards Cas A was highly polarized and aligned with the SNR magnetic field,
providing evidence that the grains are within the remnant. Similarly in the
Crab Nebula, \cite{gomez12a} found evidence of a more massive, cool
component of dust with \herschel\ with temperature $\sim 35\,$K. The ejecta
dust was distributed in the dense filaments of the plerion where molecules
are also thought to exist \citep{loh11}.  The Crab Nebula dust mass ranges
from $0.1-0.6\,M_{\odot}$ depending on the assumed grain composition and
clumping factor of the ejecta (see also \citealp{owen15}).  Sitting between
the old and young SN regimes, FIR-submm observations of SN1987A with
\herschel, and subsequently ALMA, revealed the unambiguous detection of
freshly-formed ejecta dust, finding $0.4-0.8\,M_{\odot}$ of cold dust
grains ($17-23$\,K) \citep{matsuura11,indebetouw14,matsuura15, mcCray16}.  The dust
masses derived from FIR-submm observations for these three SNRs are
therefore one to two orders of magnitude higher than that measured in young
SNe or based on NIR-mid-IR observations only.
This could be due to the
longer wavelength observations needed to detect emission from the colder,
more massive population of dust, or as proposed in \citet{gall14} (see also
\citealp{wesson15}), simply a time evolution issue: the very young SNe are
yet to build up their dust mass to the level seen in the SNRs. There is
therefore growing evidence based on individual case studies that SNe are
important contributors to the dust budget in galaxies and could be
responsible for the high dust masses seen in the early Universe.  However,
evidence for cold dust emission in SN ejecta is still based on only three
known sources.

Further evidence for SN dust formation comes from
observations of pre-solar grains: dust created in the winds of evolved
stars, supernovae \citep{clayton04} and in the interstellar medium before
our Solar System was formed.   These grains are recognized by their highly
unusual isotopic compositions relative to all other materials, with the
most abundant examples being Silicon carbide (SiC), nanodiamonds, amorphous
silicates, forsterite, enstatite, and corundum. Somes isotopic anomalies of
heavy elements found in meteorites \citep{messenger06, nguyen16} have been attributed
to dust grains expelled by SNe, these include the presolar grains: diamond,
SiC, low-density graphite, Si$_3$N$_4$ and Al$_2$O$_3$ \citep{clayton04}.

In this paper we use photometric and spectroscopic observations of the
young SNR G54.1+0.3 at IR-submm wavelengths to investigate whether dust is
present in the SNR and to determine its composition.  G54.1+0.3 is
identified as a SNR from radio non-thermal emission \citep{green85} with
strong linear polarization detected \citep{reich85}. The SNR is classified
as a Crab-like source with a flat radio spectral index of $-0.13$ and a
flux density at 1\,GHz of $0.364$\,Jy \citep{velusamy88}. G54.1+0.3 has a
small angular radius in radio and X-ray images (1.5$'$) and is thought to
be 1800-2400 yr old \citep{bocchino10}. Its distance is somewhat uncertain
with literature values ranging from $5-10$\,kpc \citep{lu02, koo08}, here
we use $6.2_{-1.2}^{+3.8}$\,kpc from \citet{leahy08}. X-ray emission was
detected with the {\it Einstein telescope}, {\it ASCA}, {\it ROSAT} and
\chandra. The latter dataset resolved the SNR into a number of distinct
X-ray structures including a ring, an outer nebula and a pulsar
\citep{lu02}. CO observations toward G54.1+0.3 show no-interaction with
clouds and instead the SNR appears to be located within a bubble
\citep{lee12, lang10}. Very high energy $\gamma$-ray emission using the
VERITAS ground-based gamma-ray observatory was detected from the young
rotation-powered pulsar in G54.1+0.3 \citep{acciari10}. \citet{temim10}
used \spitzer\ and \chandra\ observations of G54.1+0.3 to suggest that
pulsar-wind nebula is driving shocks into the expanding SN ejecta 
which is responsible for the IR morphology (see discussion 
in Section \ref{sec:sedfitting}). They
estimated dust mass in the IR shell 0.04-0.1 M$_\odot$ assuming a
forsterite grain composition but without detailed modeling of the dust
composition and properties.

In this paper,  we present 350$\mu$m (SHARC-II), 870$\mu$m (LABOCA), and
far-IR \herschel\ images of G54.1+0.3 (Section~\ref{sec:data}) in addition
to the \spitzer\ spectrum. \herschel\ images are critical to estimate the
mass of cold dust which will allow accurate dust mass, and we show detailed
dust modeling by exploring different dust grain compositions required to
match the observed spectra. We compare the dust features revealed in
G54.1+0.3 with that of the SNR Cas A. We have presented our earlier results
at meetings \citep{rho10, rho12a} and the conference of {\it F.O.E. 2015: Fifty-one
Erg}{\footnote{https://www.physics.ncsu.edu/FOE2015/}}. After we almost
complete our draft of this paper, we note the paper by \citet{temim17}. The
spectral fitting has independent methods with different grain compositions,
and our dust mass is comparable to (may be slightly smaller than) those in
\citet{temim17} (see Section \ref{sec:dustmass} for details). We specially
concentrate on pre-solar grains of SiO$_2$ to reproduce the 21$\mu$m dust
feature and use diverse choices of cold dust grains including
temperature-dependent silicate absorption coefficient and  Mg$_2$SiO$_4$,
MgSiO$_3$, Al$_2$O$_3$, FeS, carbon and Fe$_3$O$_4$. We also compare the
IRS spectra of G54.1+0.3 with various astronomical objects such as carbon
stars and protoplanetary nebulae.

\section{Observations}
\label{sec:data}

\subsection{Mid-infrared Spectroscopy with \spitzer}

\begin{figure*}
\includegraphics[width=16truecm]{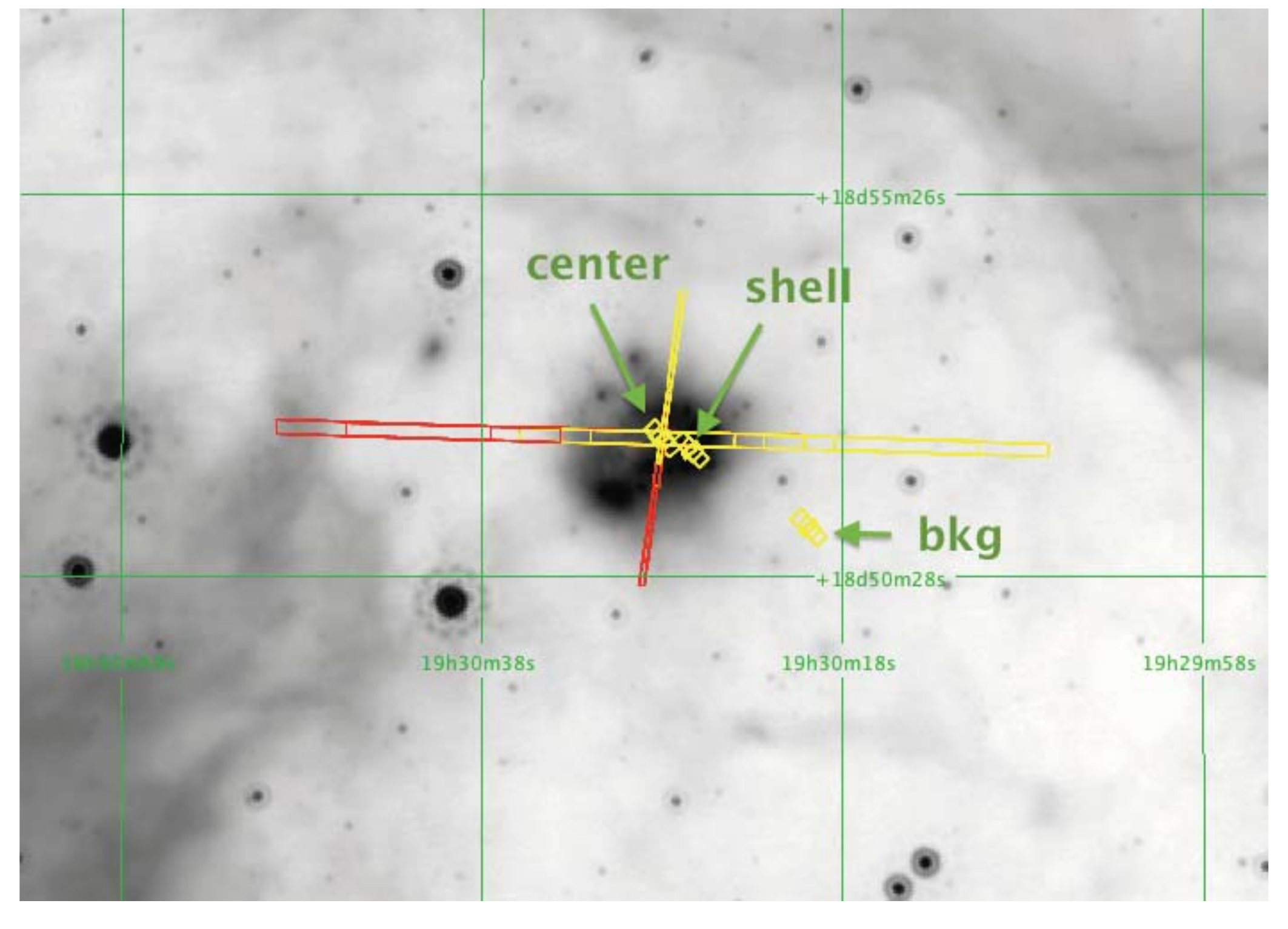}
\caption{
The \spitzer\ IRS slit superposed on a MIPS
    24$\mu$m image of the SNR G54.1+0.3.  The wide and long horizontal
    slits are the low-resolution LL1 (in red) and LL2 (in yellow)
    modules; the thin and long vertical slits are the low-resolution
    SL1 (red) and SL2 (yellow) modules. The high-resolution slits are
    the much smaller footprints shown in yellow; LH has a larger FOV
    than that of of SH. The positions of center, shell and background
    regions of high-resolution IRS spectra are labeled.}
\label{fig:g54IRSslit}
\end{figure*}

\begin{figure}

\includegraphics[width=9truecm]{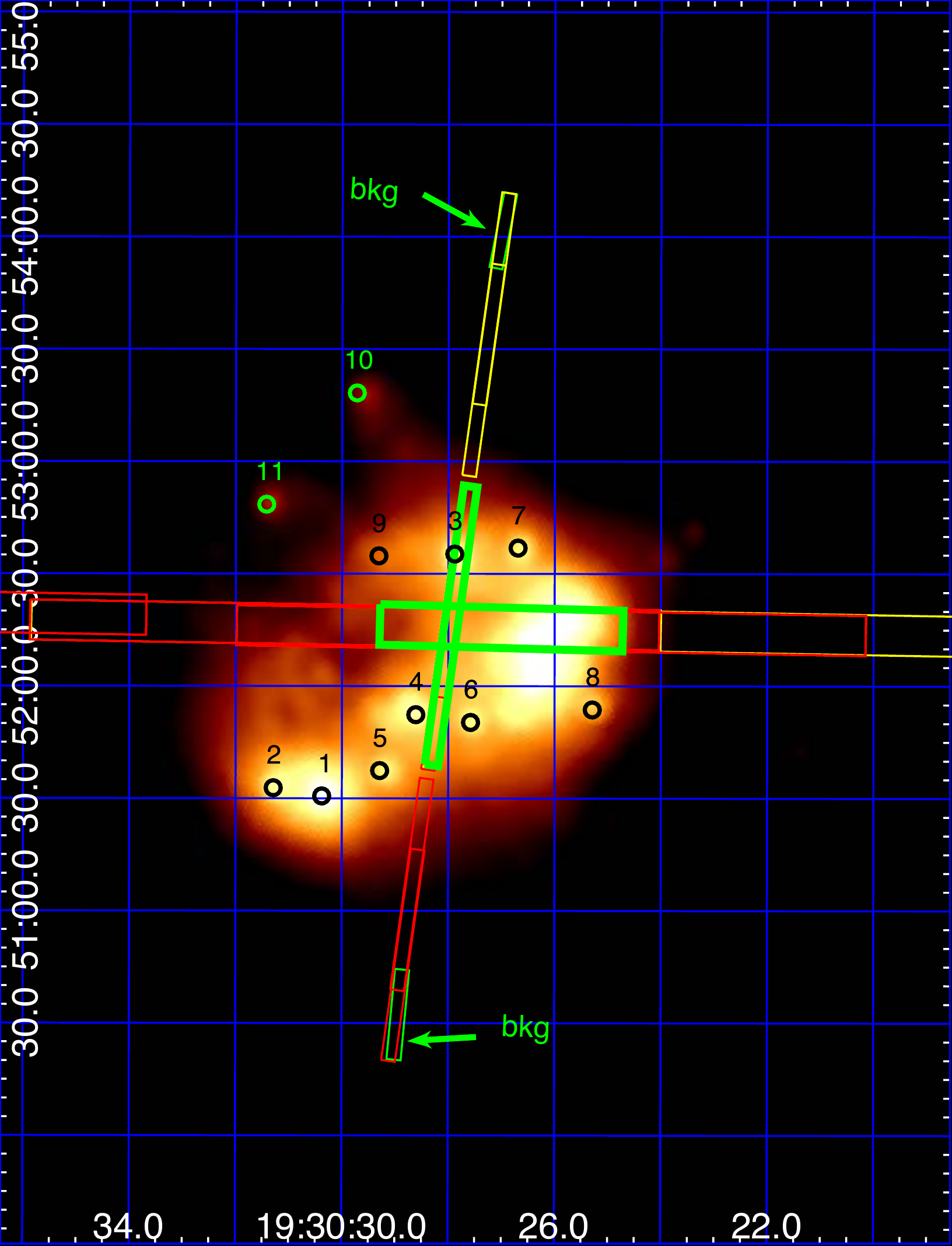}
\caption{A close up of \textcolor{blue}{the low-resolution IRS} SL1 (red, vertical), SL2
    (yellow, vertical), LL1 (red, horizontal), and LL2 (yellow,
    horizontal) are marked on the \spitzer\ 24$\mu$m image. OB stars
    identified by \protect\cite{koo08} and \protect\cite{kim13} are labeled and the
    star behind the SL slits is object 3 \protect\citep{koo08} for which
    \protect\cite{kim13} did not have spectroscopic confirmation.  The regions
    for the extracted spectra are marked in green (thick) and the
    background regions for SL are marked in green (thin; the
    background (bkg) region marked in north is for SL2 and in south is
    for SL1.}
\label{fig:g54IRSslitzoom}
\end{figure}

We used archival \spitzer\ data of G54.1+0.3 with all three instruments.
Data from the Infrared Array Camera (IRAC) was available from the GLIMPSE
survey \citep{benjamin03} and Multiband Imaging Photometer for {\it
Spitzer} (MIPS) from the MIPSGAL survey. The MIPS and IRAC observations
took place on 2005 October 6 and 24, respectively. The \spitzer\  MIPS
24$\mu$m image is shown in Figure \ref{fig:g54IRSslit}.  The bright `blob'
and diffuse emission from the remnant are seen in both images. Analysis of
the Infrared Spectrograph (IRS) dataset were previously presented by
\citet{temim10, temim17}. The IRS observations took place on 2007 November
4. Here we re-analyzed the IRS data with CUBISM \citep{smith07} in order to
directly compare with our results from Cas A \citep{rho08, arendt14, deLooze17}. We obtained
recent data products that were processed with the S18.18 version. The
spectra below 8$\mu$m were relatively low ($<$5) signal-to-noise (the rest
of spectra have up to 600 signal-to-noise) and somewhat sensitive to the
background selection. The IRS slit coverage is shown in Figure
\ref{fig:g54IRSslit} including positions at the center and western shell
(position 1 and 2 in \citet{temim10}, respectively). We extracted the IRS
SL and LL spectra (AORKEY: 23174656) from the brightest diffuse `blob' region seen toward the
western shell ($20 \times 10^{\prime \prime}$ across centered at R.A.\
$19^{\rm h} 30^{\rm m} 26.28^{\rm s}$ and Dec.\ $+18^\circ$52$^{\prime}
20.0^{\prime \prime}$, J2000). High resolution spectra with IRS were taken
toward two positions (AORKEY: 23175168); the center position is  at the same position as that
of low resolution spectrum, and the western shell position is at R.A.\
$19^{\rm h} 30^{\rm m} 26.40^{\rm s}$ and Dec.\ $+18^\circ$52$^{\prime}
07.0^{\prime \prime}$. 
High resolution off-source (background) observation was taken (AORKEY: 23174912)
toward R.A.\
$19^{\rm h} 30^{\rm m} 19.89^{\rm s}$ and Dec.\ $+18^\circ$51$^{\prime}
05.7^{\prime \prime}$. 
The final high resolution spectra for the center and shell positions were obtained after
subtracting the off-source position. The locations of the spectral coverage are shown in
Figure \ref{fig:g54IRSslitzoom}.

\begin{table*}
\caption[]{Summary of Far-IR and Submm Observations.}\label{Tobs}
\begin{center}
\begin{tabular}{llllll}
\hline \hline
Date  & Telescope & Wavelength & FWHM Beam \\ \hline
2006 Apr 18 and May 3-4 & CSO SHARC-II & 350\mic & 8.5$^{\prime \prime}$    \\
2008 Apr 27 and Jun 15  & APEX LABOCA & 870\mic &  18.6$^{\prime \prime}$   \\
2011 May 1   &\herschel\ SPIRE & 250, 350 and 500\mic & 18.1, 24.9 and 36.4$^{\prime \prime}$   \\
2012 April 9 &\herschel\ PACS  & 70, 100 and 160\mic & 6, 8 and 12$^{\prime \prime}$  \\
\hline \hline
\end{tabular}
\end{center}
\end{table*}

\subsection{Far-infrared and submm observations}

\subsubsection{\herschel}

We used archival data from the \herschel\ {\it Space Observatory}
(\citealp{pilbratt10}, hereafter \herschel) taken as part of the HiGAL
Legacy survey \citep{molinari10} for the Photodetector Array Camera (PACS;
\citealp{pogslitsch10}) 70 and 160 $\mu$m and the Spectral and Photometric
Imaging Receiver (SPIRE; \citealp{griffin10}) 250, 350 and 500 $\mu$m
images (Table~\ref{Tobs}). The observations are taken in parallel mode to
maximize survey speed and wavelength coverage and the instruments
wavelength multiplexing capabilities.
We also created a PACS 100$\mu$m image using scan-map data 
from obsids of 1342231921 and 1342231922
(OT1\_ttemim\_1). PACS maps at wavelengths of 70, 100 and
160\mic\ with a Full Width Half Maximum (FWHM) of 6, 8 and 12$^{\prime
\prime}$ and SPIRE at 250, 350 and 500\mic\ with FWHM of 18.1, 24.9 and
36.4$^{\prime \prime}$, respectively. Data are taken from the arrays
simultaneously as the spacecraft is scanned across the sky. In order to
have redundancy two passes are taken over the same region of the sky using
the other scan angle at -42.5$^{\circ}$ angle which is nearly orthogonal to
the first one and the distance between each scan in parallel mode is set by
the size of the PACS array.

The PACS and SPIRE photometric data were reduced with the \herschel\
Interactive Processing Environment (HIPE; \citealp{ott10}). For PACS, all
low-level reduction steps were applied to Level 1 using HIPE and level 1
output was imported to {\sc scanamorphos} software \citep{roussel13} to
remove effect due to thermal drift and uncorrelated noise of the individual
bolometers. For the SPIRE data, additional iterative baseline removal step
(Bendo et al. 2010) was applied and the final map was created with the
standard {\sc naive} mapper.

The flux calibration uncertainty for PACS is less than 10\%
\citep{pogslitsch10} and the expected color corrections are small compared
to the calibration errors. We therefore adopt a 10\% calibration error for
the 70 and 160\mic\ data. The SPIRE calibration methods and accuracies are
outlined by Swinyard et al. (2010) and are estimated to be 7\%. Note that
we also examined Planck archival data (e.g., \citealp{planck11}) but the
emission from G54.1+0.3 could not be positively identified due to the large
angular size of the Planck beam (4.4-32.7$^{\prime}$ from 857 to 30 GHz).

\subsubsection{SHARC-II Data Reduction}

Higher resolution observations of G54.1+0.3 at 350\mic\ were provided by
observations with SHARC-II (Table~\ref{Tobs}). SHARC-II is a 12$\times$ 32
bolometer array with field of view of $2.6^{\prime }\times 0.97^{\prime}$
on the Caltech Submillimeter Observatory in Hawaii. The Dish Surface
Optimization System (DSOS) was used throughout the run, providing a stable
beam with FWHM of 8.5$^{\prime \prime}$. The weather conditions were good
during both runs, with $\tau_{225}$ ranging from 0.04 to 0.05. 22 scans
were observed in total using the boxscan format with scanning speed of 20
scans per second with on-source integration time of 3.7\,hours. The sources
Callisto, Neptune, Arp220, G34.3 and Hebe were used for pointing, focusing
and flux correction. The focus was found to be stable within 5\,mm.
Pointing was measured to be offset by $8^{\prime \prime}$, this was
corrected in the data reduction. The data were reduced with the SHARC-II
data reduction package CRUSH version 1.6.3 \citep{kovacs06}. The background
noise in the central region is $\sim$80\,mJy\,beam$^{-1}$. The flux
uncertainties include the flux calibration uncertainty, which is typically
15\%.

\begin{figure*}
\includegraphics[width=16cm]{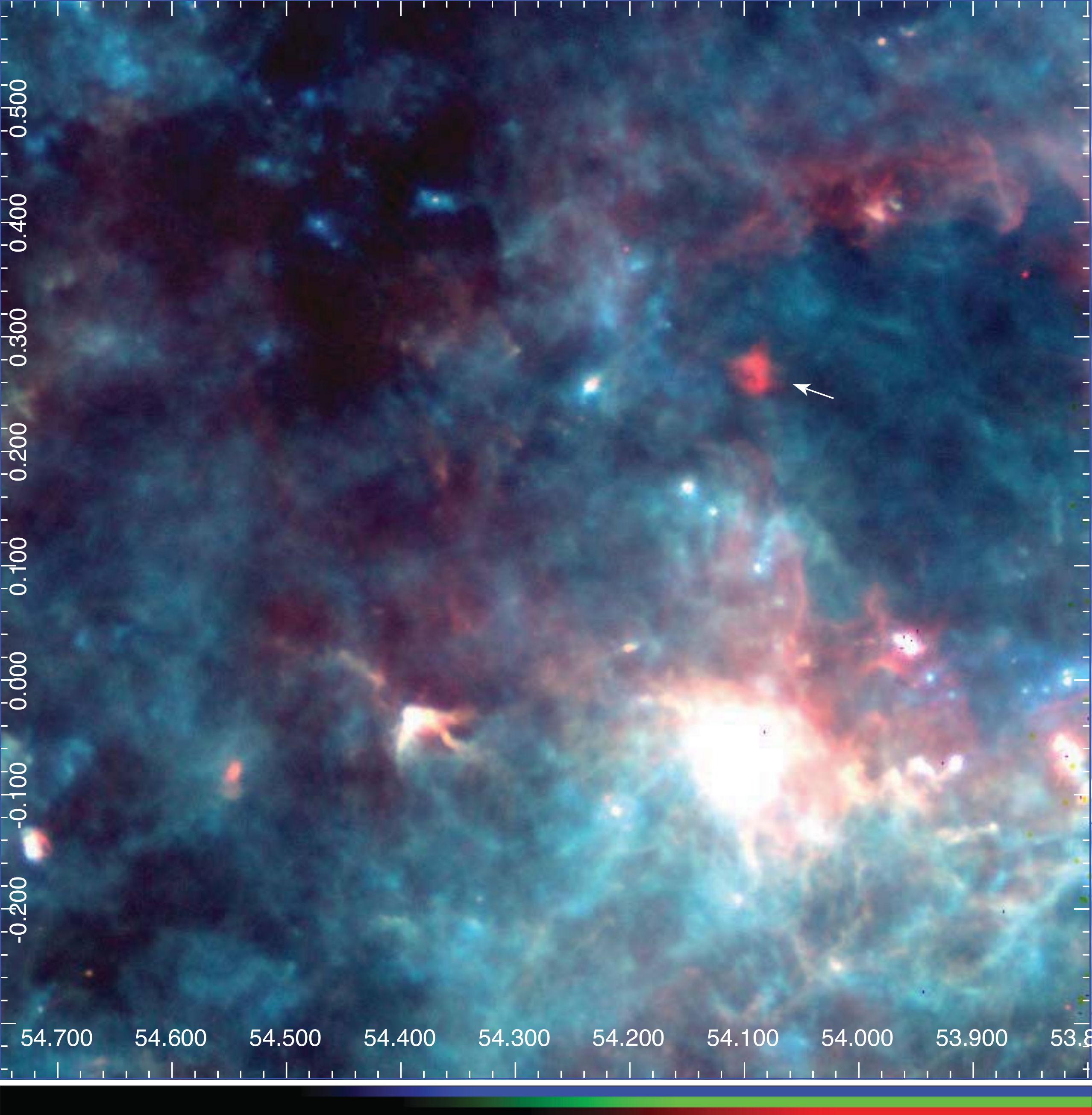}
\caption{A \herschel\ three colour image of G54.1+0.3 and its environments.
The emission at 70, 250 and 500\mic\ is shown in red,  green and blue,
respectively (warmer material appears in redder color). The SNR G54.1+0.3
is marked as an arrow and is noticeably redder (warmer) than the ISM cirrus
(in blue/green) and other emission in the immediate vicinity of the SNR.
Galactic coordinates in degree are marked, and color bar scale is 0.004 -
0.75 Jy pixel$^{-1}$ (1 pixel = 1.4$''$) in red, 0.003 - 53 Jy beam$^{-1}$
in green, and 0.06-14 Jy beam$^{-1}$ in blue.
}
\label{fig:g54herschelimagecolor}
\end{figure*}

\subsubsection{LABOCA Data Reduction}

We supplemented the \herschel\ and SHARC-II data with 870\mic\ observations
taken with the LABOCA camera \citep{siringo09}, a 295-pixel bolometer array
with beam size FWHM of 18.6$^{\prime \prime}$, located on the Atacama
Pathfinder EXperiment telescope \citep{gusten06} on Chanjantor in Chile
(Table~\ref{Tobs}).  The observations were carried out in raster spiral
mode with forty scans, each providing a fully sampled map over $8 \times
8^{\prime}$.  The total on-source integration time was 4.7\,hours.  Two
independent measurements of the optical depth, $\tau$ were obtained.  The
first method used the precipitable water vapor (PWV) levels measured every
minute along the line of sight, then scaled using the relevant atmospheric
transmission model. The PWV was low but variable, ranging from
0.5--0.8\,mm.  The second method calculated $\tau$ from skydip
measurements, where a model of the dependence of the effective sky
temperature on elevation were fitted to determine the zenith opacity. The
two skydips taken before and after the on-source scans were both well
fitted by the model, with $\tau$ between 0.16-0.19.  These values were, on
average, 30\% lower than those estimated from the PWV measurements.  A
linear combination of the two methods, and a final comparison with the
calibrator models (e.g. \citealp{siringo09}), was used to determine the
values used in the data reduction.

The data were reduced using the BoA (BOlometer array Analysis software)
package.  The focus was checked on observations of Jupiter and G34.3 and
were stable within $\pm$0.2\,mm. Pointing observations of Jupiter were
within $2^{\prime \prime}$ in azimuth and elevation. Bad and noisy pixels
were flagged. The data were de-spiked and correlated noise was removed
\citep[removing a large fraction of the sky noise][]{schuller09}.  The
reduction was optimized for the recovery of weak sources.  All of the scans
were coadded (weighted by rms$^{-2}$) to create the final map. After a
first iteration of the reduction, the source map was used to flag bright
sources and the data were reduced again. This was efficient at removing
negative artifacts which appeared around the bright sources in the first
iteration and led to a more stable background noise level in the central
region ($\sim$10\,mJy\,beam$^{-1}$). The calibrators Jupiter, G34.3,
J1925+211 and J1751+097 were used to calibrate the map. The total
uncertainty in the calibration, including uncertainties in the calibrator
model is $\sim$12\%.

\begin{figure*}
\includegraphics[width=18.2truecm]{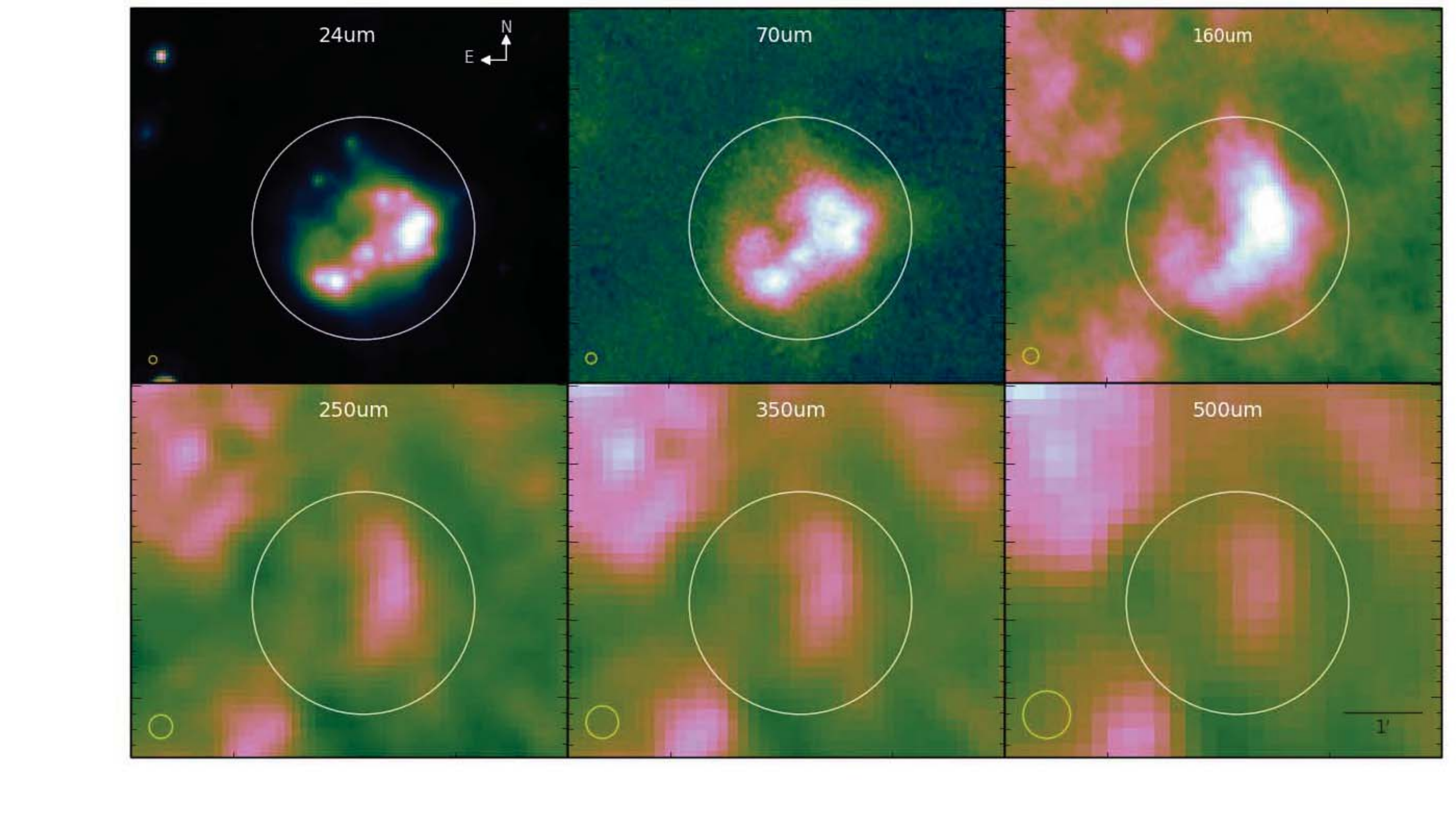}
\caption{Comparison of \spitzer\ and \herschel\ images of G54.1+0.3. {\it
Top:} \spitzer\ 24\mic\  (left), \herschel\ PACS 70\mic\ (middle) and
100\mic\ (right). {\it Bottom:} \herschel\ SPIRE 250\mic\ (right), 350\mic\
(middle) and 500\mic\ (right). Beam sizes are shown in the lower right
corner, the large circle has a radius 86$^{\prime \prime}$ centered on R.A.\ $19^{\rm h}
30^{\rm m} 28.42^{\rm s}$ and Dec.\ $+18^\circ$52$^{\prime}6.6^{\prime
\prime}$ (J2000).  The dust
emission in the PACS and SPIRE images is distributed in a shell-like
morphology, similar to that seen in the \spitzer\ 24\mic\ image (tracing
warm dust). The shell-like morphology in the PACS images is elongated from
left to right, but the emission in the SPIRE images (tracing colder dust)
is elongated from north to south with a structure similar to the emission
seen in the \spitzer\ 8\mic\ image.}
\label{fig:g54herschelimagenew}
\end{figure*}

\begin{figure*}
\includegraphics[width=14truecm]{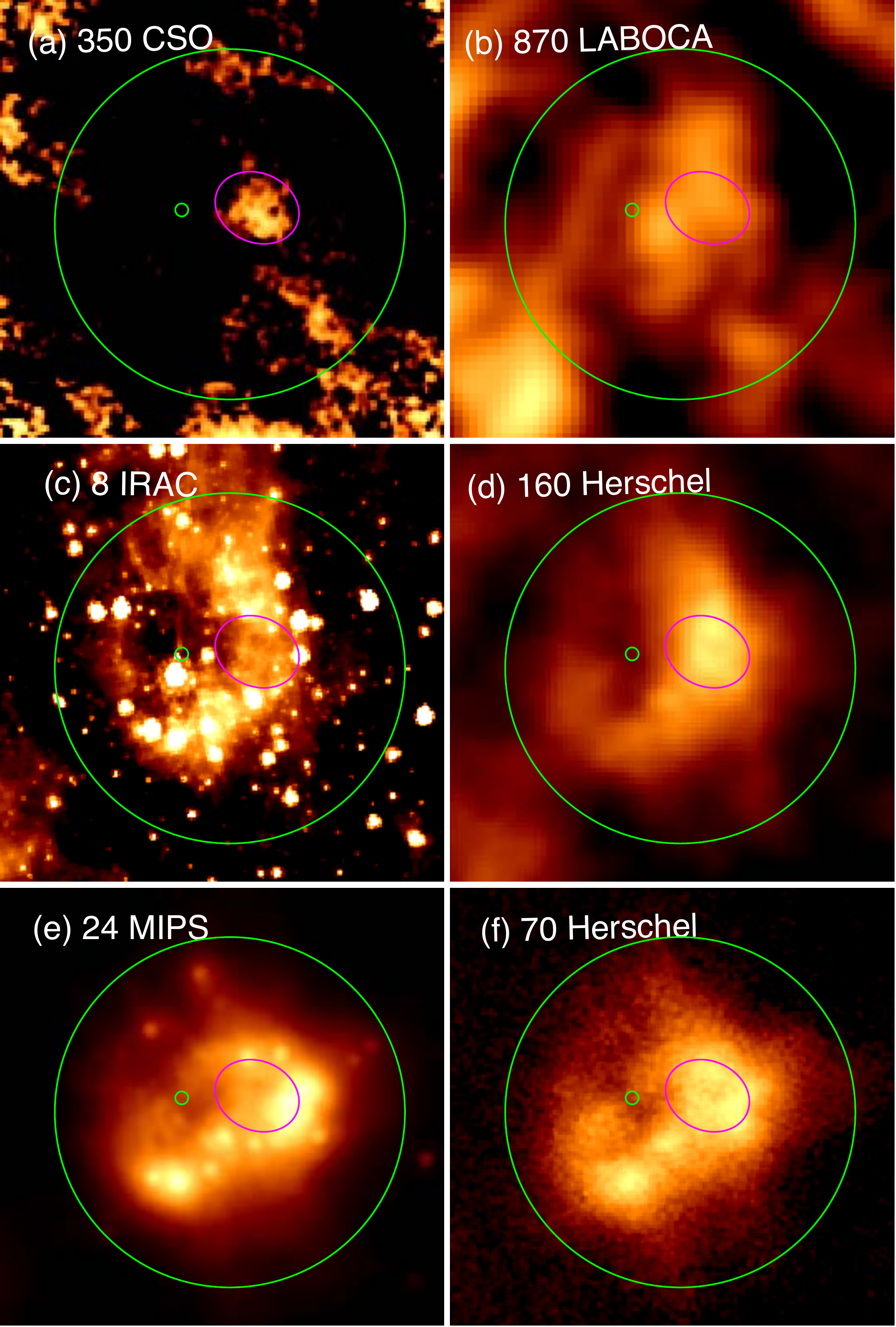}
\caption{High resolution submm and mid-IR images of G54.1+0.3. {\it (a)}
CSO SHARC-II at 350$\mu$m (the color scale ranges  0.05 -- 0.33 Jy
beam$^{-1}$).  {\it (b)} LABOCA at 870$\mu$m (7$\times$10$^{-5}$
-- 0.36 Jy beam$^{-1}$).  {\it (c)} \spitzer\ IRAC 8$\mu$m
(15-21 MJy sr$^{-1}$), and {\it (d)} {\it Herschel} PACS
160$\mu$m (0.017 -- 0.045 Jy pixel$^{-1}$).
 {\it (e)} \spitzer\ IRAC 24$\mu$m
(53-224 MJy sr$^{-1}$), and {\it (f)} {\it Herschel} PACS
70$\mu$m (0.007 -- 0.028 Jy pixel$^{-1}$). 
The SHARC-II submm image (note
that this observation was taken earlier than those from Herschel as listed
in Table 1) reveals submm emission from cold dust in G54.1+0.3 (marked as a
purple circle), which coincides the bright part of the Herschel 350$\mu$m
image in higher spatial resolution.  The small
and large circles (in green) mark the locations of the pulsar J1930+1832
and the entire SNR, respectively \citep{lang10}.}
\label{fig:g54submm}
\end{figure*}

\begin{figure}
\includegraphics[width=8.2truecm]{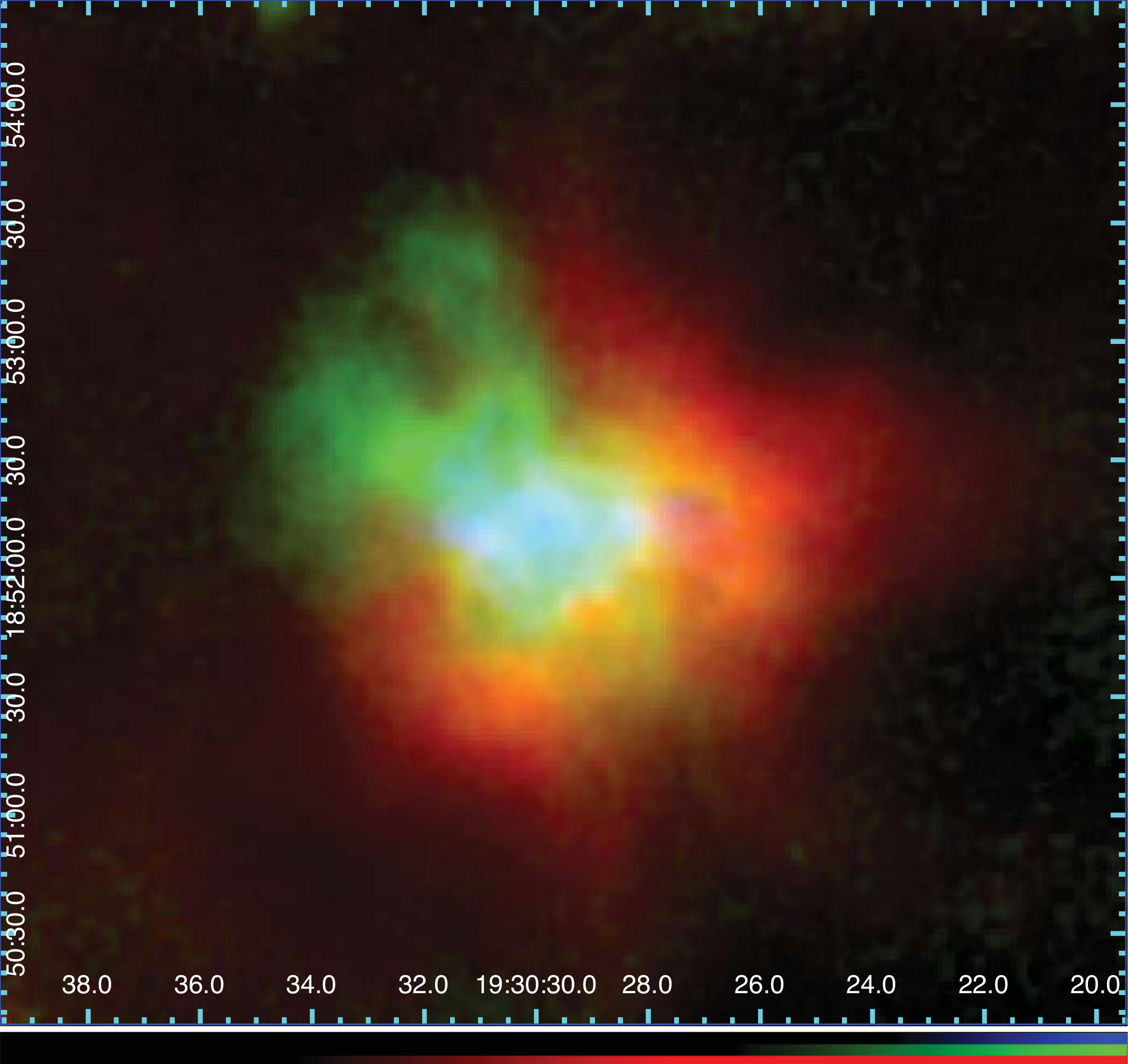}
\caption{
 Three color image of G54.1+0.3 with blue, green and red colour
indicating X-ray, radio and FIR ({\it Herschel} PACS at 70$\mu$m) emission
respectively. The \herschel\ image shows a torus-like shell emission
(15$^{\circ}$ tilted from the horizon) and consistent with the 24\mic\
emission. The radio pulsar nebula is perpendicular to this torus structure.
The \herschel\ images (see Figure \ref{fig:g54herschelimagenew}) also show an
emission feature extended north of this shell on the western side of the remnant.
The north is up and the west is right. 
Color bar scale is 0.62 - 13.20 $\times$10$^{-6}$ photons cm$^{-2}$ s$^{-1}$ in blue, 0.15 - 1.75 MJy sr$^{-1}$ in green, and 0.004 -
0.75 Jy pixel$^{-1}$ (1 pixel = 1.4$''$) in red.}
\label{fig:g54multi}
\end{figure}

\begin{figure*}
\includegraphics{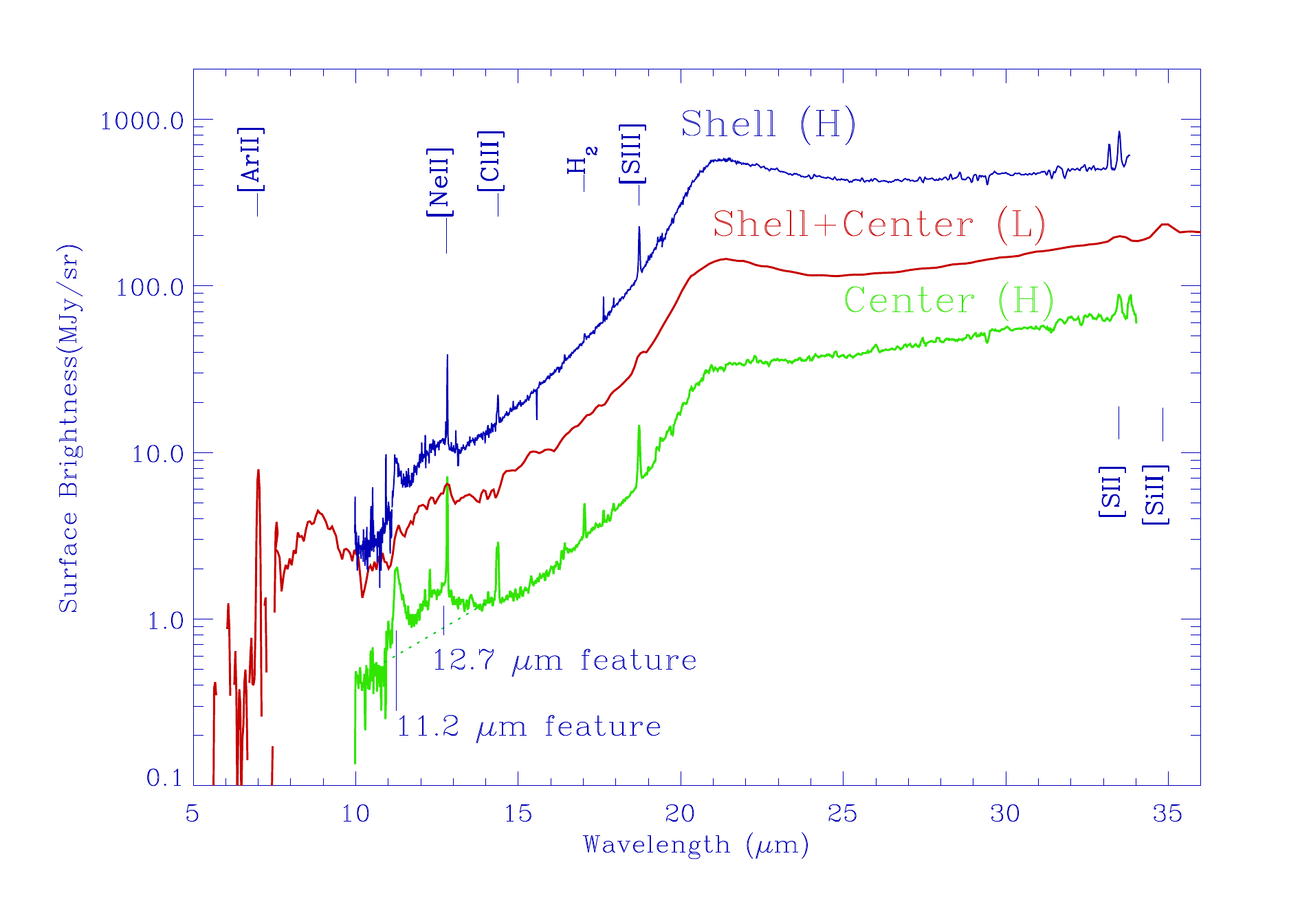}
\caption[]{\spitzer\
IRS spectra of G54.1+0.3. The red curve is the low-res spectrum and the blue
is from the high-res spectrum from the shell component of the SNR (the
bright `blob' emission in the southwestern shell). The green curve is the
high res spectra measured in the central region. The line at 14.35$\mu$m
shows multiple (high velocity) components in emission as listed in Table
\ref{tab:Tg54irslines}.
}
\label{fig:g54irsspec}
\end{figure*}

\section{Results}
\label{sec:results}

\subsection{G54.1+0.3 in Many Colours}
\label{sec:multiwavelengthview}

Here we discuss the appearance of the SNR in the IR-submm. A three color
FIR-submm \herschel\ image (70$+$350$+$500\mic) in
Figure~\ref{fig:g54herschelimagecolor} shows the dust structures in the
local area. The SNR itself can be seen at the center of this image and is
noticeably redder (hotter) than the immediate surroundings, suggesting that the
70\mic\ emission is indeed associated with the remnant. H~II regions and
young stellar objects are instead bright in all three bands, and appear
white in this image, with interstellar cirrus from unrelated foreground and
background dust emission appearing blue.

Zooming in on the remnant, the MIR images (Figure \ref{fig:g54submm}),
show bright emission from a shell-like structure (hereafter the shell)
located at the southern end of the radio emission, with peak brightness in
the western side. We see IR emission extending north from the western peak
of the shell. In Figure~\ref{fig:g54herschelimagenew}, we next compare the
\spitzer\ MIR data to the FIR images.  At long wavelengths
($\lambda>100$\mic), we see a slight change to the distribution of the IR
emission: the eastern part of the shell begins to fade in brightness and
the northern elongated structure becomes more prominent. This feature is
particularly obvious when we resolve the \herschel\ emission in the higher
resolution LABOCA 870\mic\ image (Figure~\ref{fig:g54submm}), though the
peak emission in the LABOCA image is slightly offset towards the pulsar
location compared to the 24 and 70\mic\ emission.  The elongated northern
structure seen in the \herschel\ and LABOCA data is coincident with the
same structure seen in the IRAC 8\mic\ image (Figure~\ref{fig:g54submm}).
The SHARC-II 350\mic\ image provides us with a means to resolve out the
submm emission seen with \herschel\ at 350\mic: here we see a smaller
structure coincident with the brightest peaks in the 24, 70 and 160\mic\
images (in the western part of the shell). The detection of submm and
far-IR detection from G54.1+0.3 and its coincidence with the IRAC 8\mic\
emission (western peak and elongated north west structure) and the
\spitzer\ 24\mic\ (shell structure) hints that the FIR emission could
originate from cold dust associated with the SNR (we return to this in
Section~\ref{sec:sedfitting}). The CSO SHARCII submm image
at 350$\mu$m has a factor of 3 higher spatial resolution compared to the
SPIRE image at 350$\mu$m, and the SHARCII image shows small scale
structures that resemble those in the 8\mic\ image (see Figures
\ref{fig:g54submm}a and \ref{fig:g54submm}c). 

In Figure~\ref{fig:g54multi} we compare the FIR emission with
the well studied X-ray and radio data for G54.1+0.3 using archival {\it
Chandra} and Very Large Array (VLA, 4.45\,GHz) observations. In X-ray 
we see the pulsar at the center and the pulsar wind nebula
extending east - west (e.g \citealp{lu02}). In radio, we
see many different structures including lobes, filaments, and diffuse
emission (see \citealp{lang10} for more information). The radio emission in
general extends perpendicular to the X-ray with a narrow `waist' region at
the center coincident with the X-ray diffuse structure. We also see
asymmetry in the radio emission with the remnant appearing brighter in the
south. The `double-pronged' feature in the north appears jet-like. (We note
that we see no radio feature coincident with the pulsar, \citealp{lang10}.)
When we compare with the FIR emission (using \herschel\ 70\mic, bottom
left, red) we see that the bright FIR shell-like emission is distributed to
the south of the X-ray emission and on the southern edge of the diffuse
radio that traces the pulsar wind.  Figures~\ref{fig:g54herschelimagenew} 
and \ref{fig:g54multi}
show clearly that the FIR/submm structure that extends north from the
shell on the western side also encompasses the X-ray and radio at the
center, tracing the outer edge. \citet{lang10} suggested that the IR
shell-like emission was due to the interaction of the PWN with an
interstellar cloud. In contrast, \citet{temim10, temim17} suggested that
the PWN is driving shocks into an expanding SN ejecta. The structure seen
in  Figure~\ref{fig:g54multi} is remarkably similar to that
of the Crab Nebula where the FIR emission also surrounded the inner radio
and X-ray emission \citep{gomez12b}.  In that work, the FIR emission was
shown to originate from dust formed in the SN ejecta and not due to the
interaction with circumstellar or interstellar material despite being
brighter in the outer regions of the X-ray and radio emission. They
suggested that the Crab Nebula does not have a forward shock such that the
pulsar wind nebula is able to expand into the freely expanding SN ejecta
\citep{hester08,smith13}. As well as a low interstellar density surrounding
the Crab, this can be explained if the Crab was the result of a low-energy
Type IIn-P event from a relatively low mass star \citep{smith13}. In this
scenario the SN ejecta interacts with a thin shell of circumstellar
material and the expansion of the pulsar wind nebula accelerates and
fragments the Crab filaments. Given that the PWNs within G54.1+0.3 and the
Crab Nebula have shown to be similar \citep{kim13}, we now posit in this
work that a similar mechanism (dust formed in the ejecta material) could be
responsible for the IR-submm appearance of G54.1+0.3. 
Based on spectral ejecta lines and dust features similar to that of Cas A,
we interpret the IR emission is from ejecta and associated dust. However,
since the spectral coverage is limited, we can not rule out contribution
from interactions of PWN with interstellar clouds. We will return to
this in Sections ~\ref{sec:spitzerirs} and \ref{sec:sedfitting} for
details.

\subsection{Ejecta Lines and Dust from MIR spectroscopy}
\label{sec:spitzerirs}

\begin{figure}
\includegraphics[width=8.5truecm]{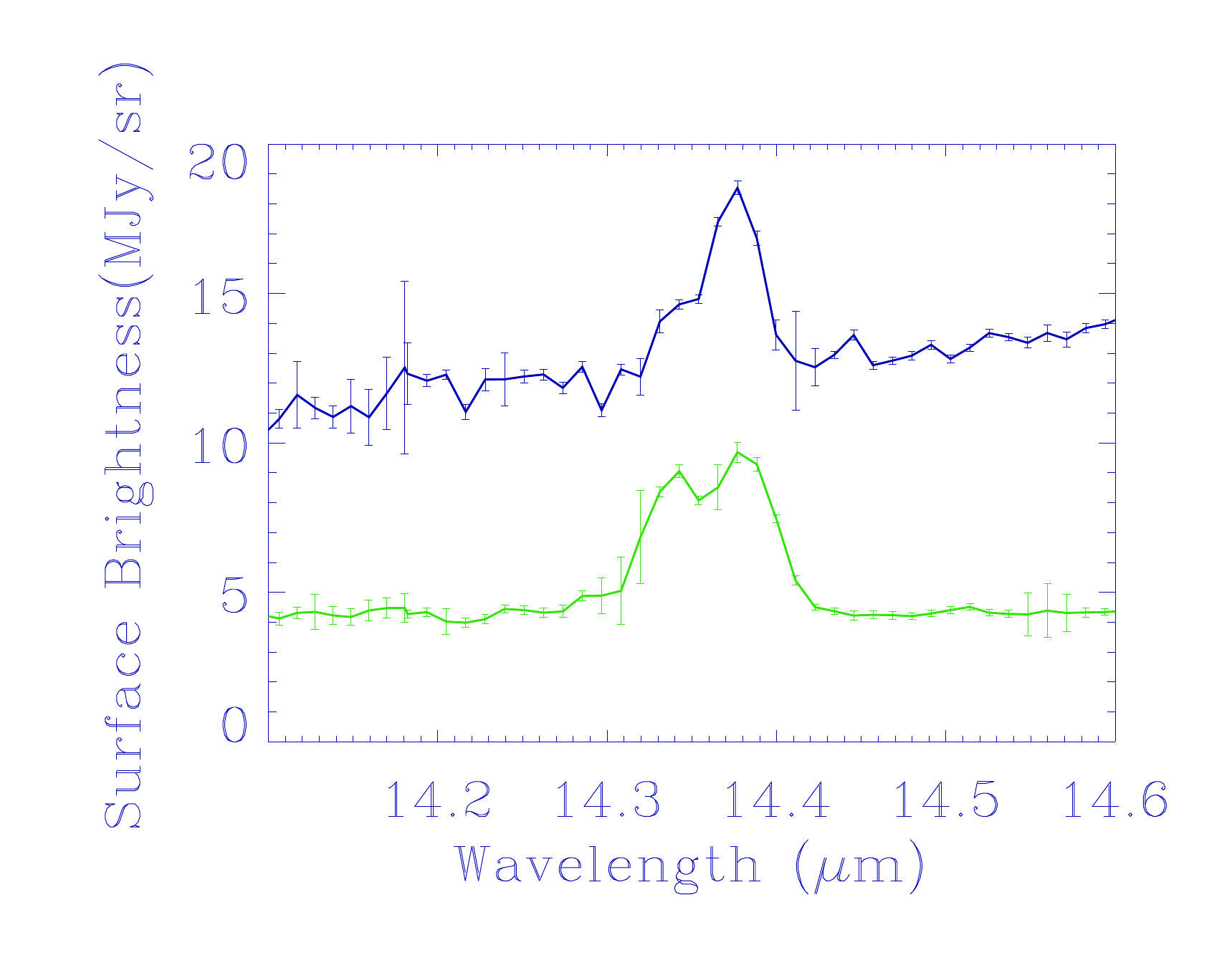}
\caption[]{Line width of the [Cl~II] emission seen at 14.36\mic. The spectra from the
shell and central regions are shown in blue and green, respectively. The
line shows multiple (high velocity) components in emission (the
two-component gaussian fits listed in Table \ref{tab:Tg54irslines}). }
\label{fig:g54irsspec14}
\end{figure}

\begin{figure}
\includegraphics[width=8.5truecm]{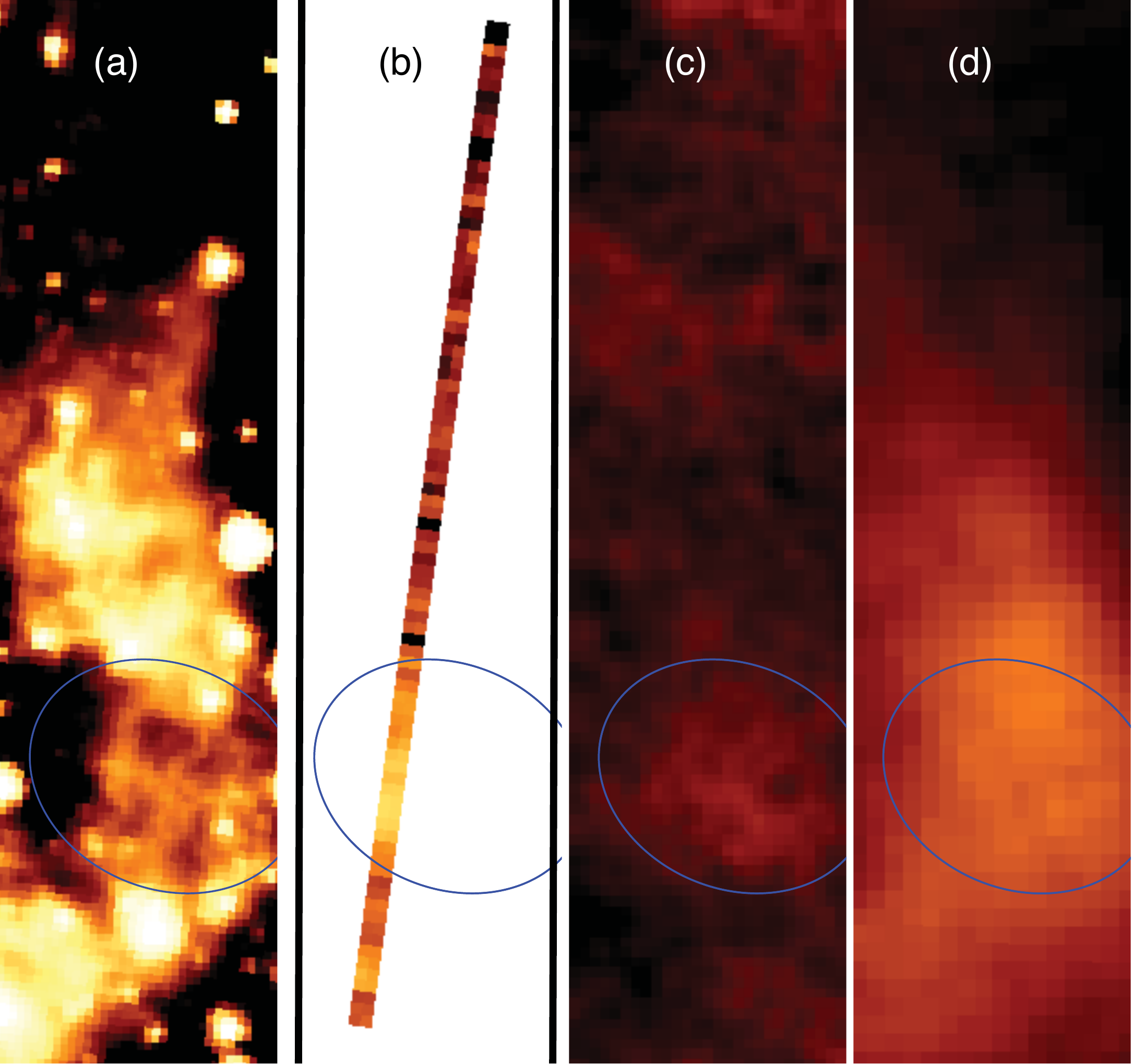}
\caption{Comparison of Ar line and continuum maps: 
(a) \spitzer\ IRAC 8\mic\ image, (b) [Ar~II] at 6.99\mic\ image 
produced using the IRS data (in staring mode;i.e., a IRS slit) observation 
after the continuum is subtracted, 
(c) CSO SHARCII 350\mic\ image, and (d) \herschel\ 160\mic\ image. 
The SHARCII 350\mic\ emitting region is marked as an ellipse. 
The [Ar~II] image shows the ejecta morphology similar to the IRAC 
8\mic\ image and SHARCII 350\mic\ dust emission within the elliptical
region, and [Ar~II] ejecta emission is correlated to the dust emission.
 Note that SHARCII 350\mic\ image (c) has a factor of 1.5 higher spatial 
 resolution to that of the Herschel 160\mic\ image (d).  
}
\label{fig:g54arslitimage}
\end{figure}

\begin{figure}
\includegraphics[height=6.7truecm]{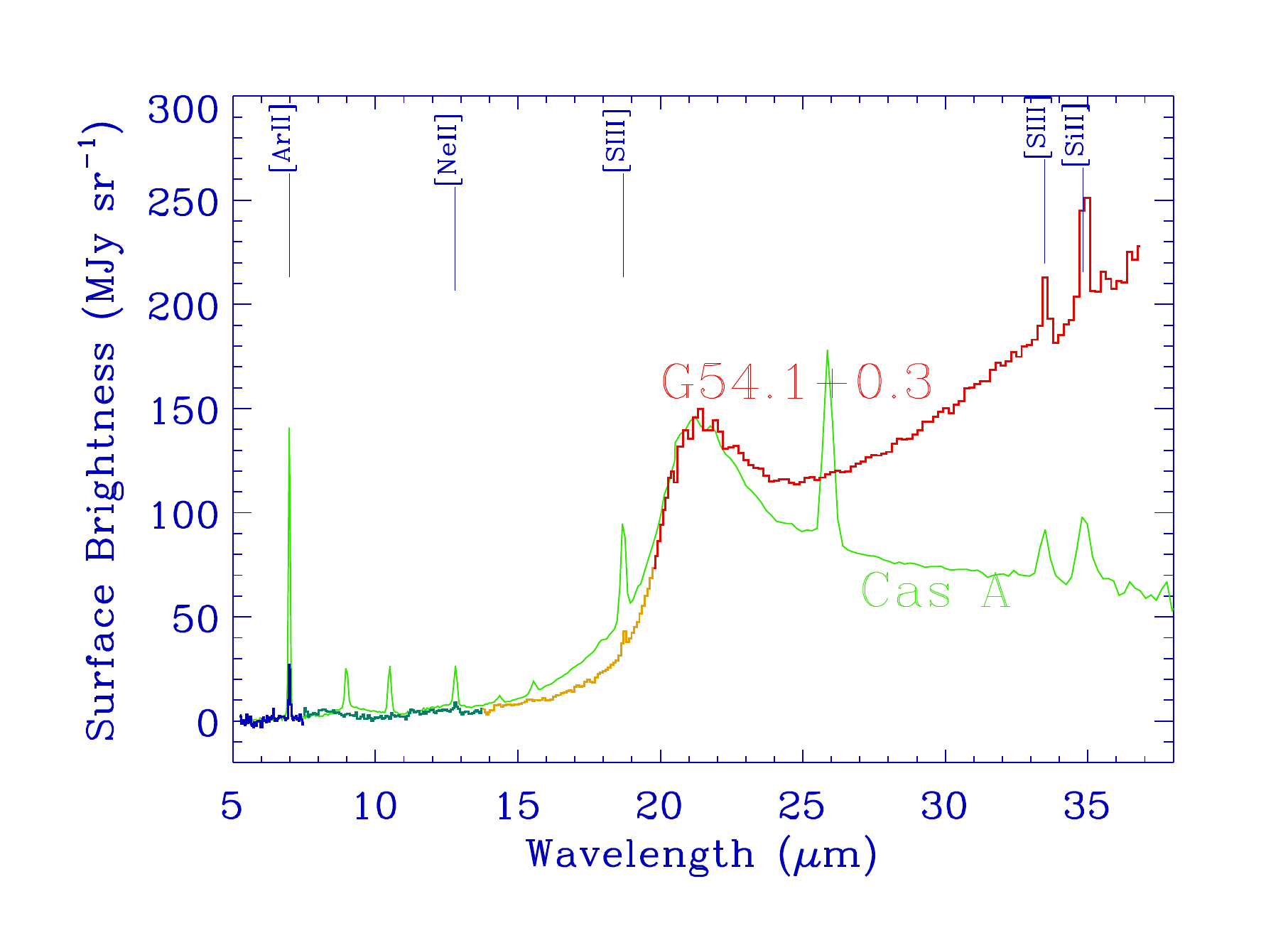}
\includegraphics[height=6.7truecm]{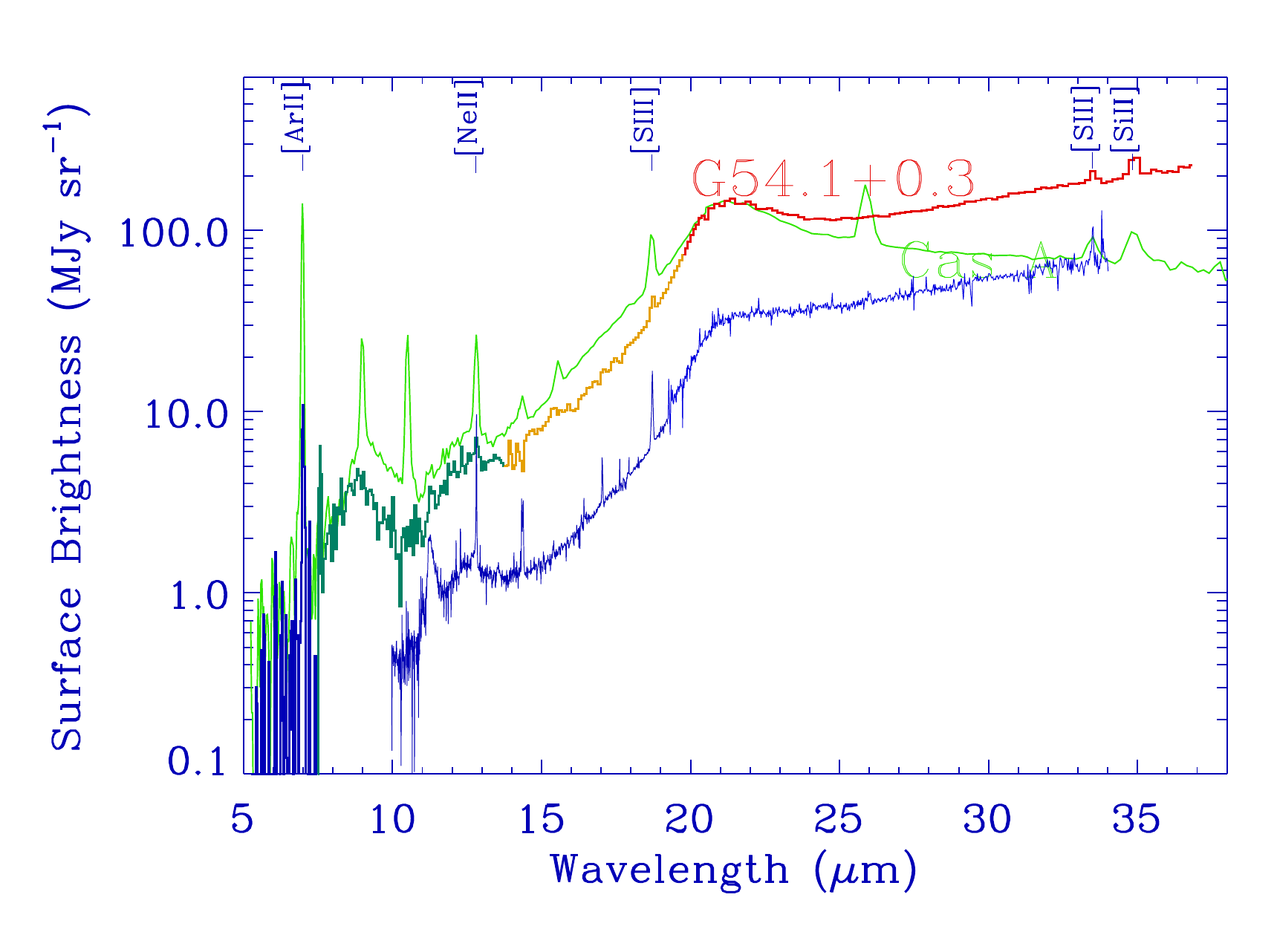}
\caption{Comparing the \spitzer\ IRS MIR spectrum for G54.1+0.3 with Cas A.
The 21$\mu$m dust feature of G54.1+0.3 is similar to that of Cas A,
showing that the feature is unique dust originated from supernovae.
The spectra shown on log-scale highlights the dust feature
seen at $\sim$12.7\mic\ (between 10 and 14\mic) in the low res spectrum (SL
in green, LL2 in yellow and LL1 in red) and the `sharper' dust feature at
11\mic\ seen in the high resolution spectrum (in blue) of G54.1+0.3.}
\label{fig:g54comparecasa}
\end{figure}

\subsubsection{Supernova Ejecta Lines}

\begin{table*}
\caption[]{Observed Spectral Line Brightness from \spitzer\ IRS spectrum. $^a$ - The spectral lines are fit with two (2) or one (1) Gaussian component(s) (Figure~\ref{fig:g54irsspec}).} \label{tab:Tg54irslines}
\begin{center}
\begin{tabular}{lrrllrrr}
\hline \hline
Region & Wavelength & Line & FWHM     &  Surface Brightness &De-reddened S.B. & Velocity & Shift \\
   &($\mu$m)   &    & $\mu$m      &  (\sbunit)    &  (\sbunit)   & (\kms)  & (\kms) \\ \hline
shell & 6.9911$\pm$    0.0012& [Ar II] &   0.0682$\pm$    0.0033& 1.30($\pm$0.07)E-04 & 1.43($\pm$0.08)E-04
 & ...& ... \\
(lowres)&   12.8231$\pm$    0.0065& [Ne II] &0.1481$\pm$0.0169& 6.49($\pm$0.77)E-06& 7.68($\pm$0.90)E-06& ... & ... \\
&   18.7338$\pm$    0.0021& [S III] &   0.1302$\pm$    0.0058& 9.27($\pm$0.41)E-6& 1.11($\pm$0.05)E-05 & ... & ...\\
&   33.5004$\pm$    0.0062& [S II] &   0.2599$\pm$    0.0100& 2.12($\pm$0.81)E-05& 2.31($\pm$0.09)E-05 &... & ...\\
&   34.8920$\pm$    0.0023& [Si II] &   0.3055$\pm$    0.0086& 5.13($\pm$0.14)E-05& 5.55($\pm$0.15)E-05& ... & ...\\
\hline
shell (2$^a$) &14.3374$\pm$0.0051 & [Cl II]  & 0.0173$\pm$0.0213 & 6.34($\pm$0.78)E-07 &7.13($\pm$0.88)E-07 &384$\pm$447 & -634$\pm$107 \\
(hires) &14.3748$\pm$0.0028 & & 0.0331$\pm$0.0074 & 3.18($\pm$0.71)E-06&3.58($\pm$0.80)E-06 & 680$\pm$153 & 129$\pm$58 \\
shell (1$^a$) & 14.3729$\pm$0.0066 & [Cl II]  &  0.0399$\pm$0.0167 &3.51($\pm$1.46)E-06 & 3.95($\pm$1.64)E-06 & 832$\pm$348 & 106$\pm$137  \\
center (2$^a$)&14.3347$\pm$0.0013 & [Cl II]  &0.0185$\pm$0.0039 & 2.08($\pm$0.44)E-06 &
2.34($\pm$0.49)E-06 &386$\pm$081 &-690$\pm$30 \\
(hires)&14.3801$\pm$0.0022 &  &0.0371$\pm$0.0057 & 3.70($\pm$0.57)E-06 &4.16($\pm$0.64)E-06 & 777$\pm$118 &+254$\pm$46  \\

\hline
\end{tabular}
\end{center}
\renewcommand{\baselinestretch}{0.8}
\end{table*}

The \spitzer\ IRS spectra include a set of low-res spectra (L, resolving
power of 65-120) towards the southwestern shell and two sets of high-res
(H, resolving power of 600) spectra toward the center and the southwestern
shell (Figure \ref{fig:g54irsspec}). The low-res IRS spectra show bright
[Ar~II] emission at 6.9$\mu$m, [Ne~II] at 12.8$\mu$m, [S~III] at 18.7 and
33.5$\mu$m, and [Si~II] at 34.89$\mu$m.

The surface brightnesses of the low-res lines are corrected for extinction
using $N_H = 1.6 \times 10^{22}\,\rm cm^{-2}$ from X-ray measurements
\citep{lu02} (see Table~\ref{tab:Tg54irslines}). The line brightnesses of high-res spectra were
listed in \citet{temim10}. Our estimate shows that the
emission seen in the IRAC 8\mic\ image (see Figure~\ref{fig:g54submm}) is
composed of Ar ejecta at 20 percent and 9$\mu$m broad dust feature at 80 percent using
the IRS low-res spectrum after convolving the IRAC transmission curve (see
Figure 1 of Reach et al. 2016). The 9$\mu$m broad dust feature is fit with
silica dust and is associated with the SN ejecta (see Section 4.2 for
details).

We also detect the [Cl~II] line at $\sim$14\mic. The rest wavelength of
this line (14.3678\mic) is very close to [Ne~V] (14.3217\mic), though we do
not detect any emission from [Ne~V] in the expected 24.3\mic\ line. The
line profile for [Cl~II] is shown in Figure \ref{fig:g54irsspec14}.
Although it is not resolved in the low-res spectrum, the high-res IRS
spectrum resolves the line into two velocity components seen in both the
shell and central regions of the SNR. 
We fit them using two gaussians and the central
velocities and widths with their errors are given in Table
\ref{tab:Tg54irslines}; both spectra show that the two knots are located at
the central velocity of  $-650\rm \,km\,s^{-1}$ (located at the front side
of the SNR) and $\sim$200$\rm \,km\,s^{-1}$ (located at the backside of the
SNR), and their widths are $\sim$380 \kms\ and $\sim$700 \kms,
respectively, which indicates that they are high velocity ejecta knots.
This is different from \citet{temim10} who
measured the velocity using high-res spectra and found a single gaussian
component was adequate.  
The low-res spectra used here covers regions that
were not covered in their analysis and demonstrate different line profiles
requiring multiple-components.  Combined, this work indicates there are at
least two ejecta clumps of the fast-moving ejecta material existing in
different places along the line of sight.

Interesting dust features are seen at 21 and 9$\mu$m; the former is a
smooth peak increasing in flux up to 21$\mu$m and decreasing beyond this.
Another dust feature appears as a bump at 11.3$\mu$m (10-13$\mu$m). The
high-res spectra shows that the feature has additional peak at 
11.2$\mu$m (see next Section). The IRS spectrum
(Figure~\ref{fig:g54irsspec}) indicates that {\it Spitzer} IRAC 8$\mu$m map
in Figure \ref{fig:g54submm} is from a combination of the Ar eject map and
the 9$\mu$m dust. 

Understanding the correlation between dust and Ar
  ejecta requires spectral mapping to cover the entire SNR as it was
  done for Cas A \citep{rho08, smith09}. Unfortunately, such data are
  not available for G54.1+0.3. We still generated the Ar image shown
  in Figure \ref{fig:g54arslitimage}b by using the IRS staring mode
  (one slit) after subtracting the continuum by spectral fitting. The
  [Ar~II] image is compared to the \spitzer\ 8\mic\,, CSO SHARCII
  image at 350\mic\ and Herschel 160\mic\ image. The [Ar~II] ejecta
  shows a morphology similar to the SHARCII 350$\mu$m dust emission
  (see Figure \ref{fig:g54arslitimage}), indicating that the Ar ejecta
  are associated with the dust. The IRAC 8\mic\ image shows
  small-scale structures similar to those in the SHARCII image at
  350\mic\ whereas the image shows large-scale structures similar to
  those in the {\it Herschel} 160$\mu$m cold dust map, in particular,
  in the northwestern shell. The IRAC 8\mic\ image seems to contain
  not only Ar ejecta emission but also additional continuum emission
  which may be associated with a combination of cold and warm dust.  

There is additional supporting evidence that the Ar
  ejecta of G54.1+0.3 might be associated with dust based on the
  analogy of the SNR G54.1+0.3 to Cas A. Both G54.1+0.3 and Cas A show
  a 21$\mu$m dust feature (see \S\ref{sec:21peak} for details, see
  Figure \ref{fig:g54irsspec}), and the map of Ar ejecta in Cas A is
  remarkably similar to the 21$\mu$m dust map where dust is formed in
  the SN ejecta \citep[see Figure 2 of ][]{rho08}.  The element Ar is
  one of the oxygen burning products at the inner-oxygen and S-Si
  layers, with the 21\mic-peak dust forming around these
  nucleosynthetic layers (e.g., \citealp{woosley95}, \citealp[see more
    discussion in][]{rho08, nozawa03}).

In the Crab Nebula, the dust is found to be associated with the ejecta
\citep[][and references therein]{gomez12}. We suggest that in SNR
G54.1+0.3, the dust is also associated with the ejecta. The analogy of the
SNR G54.1+0.3 to the Crab Nebula, where the infrared and optical/infrared
shells show overabundant ejecta knots, and the high resolution images show
almost one-to-one correspondence between ejecta and dust knots with
prominent finger-shaped knots caused by Rayleigh-Taylor instabilities
\citep[more discussion in Section \ref{sec:sedfitting};][]{hester08,
jun98}, suggests that the ejecta of G54.1+0.3 are likely associated with
the dust. Hopefully, future spectral mapping such as with JWST will further
confirm the correlation between the ejecta and dust for the entire SNR of
G54.1+0.3.

\subsubsection{21\mic\ dust feature: a dust twin of Cas A}
\label{sec:21peak}

\begin{figure*}
\includegraphics[height=11.2truecm]{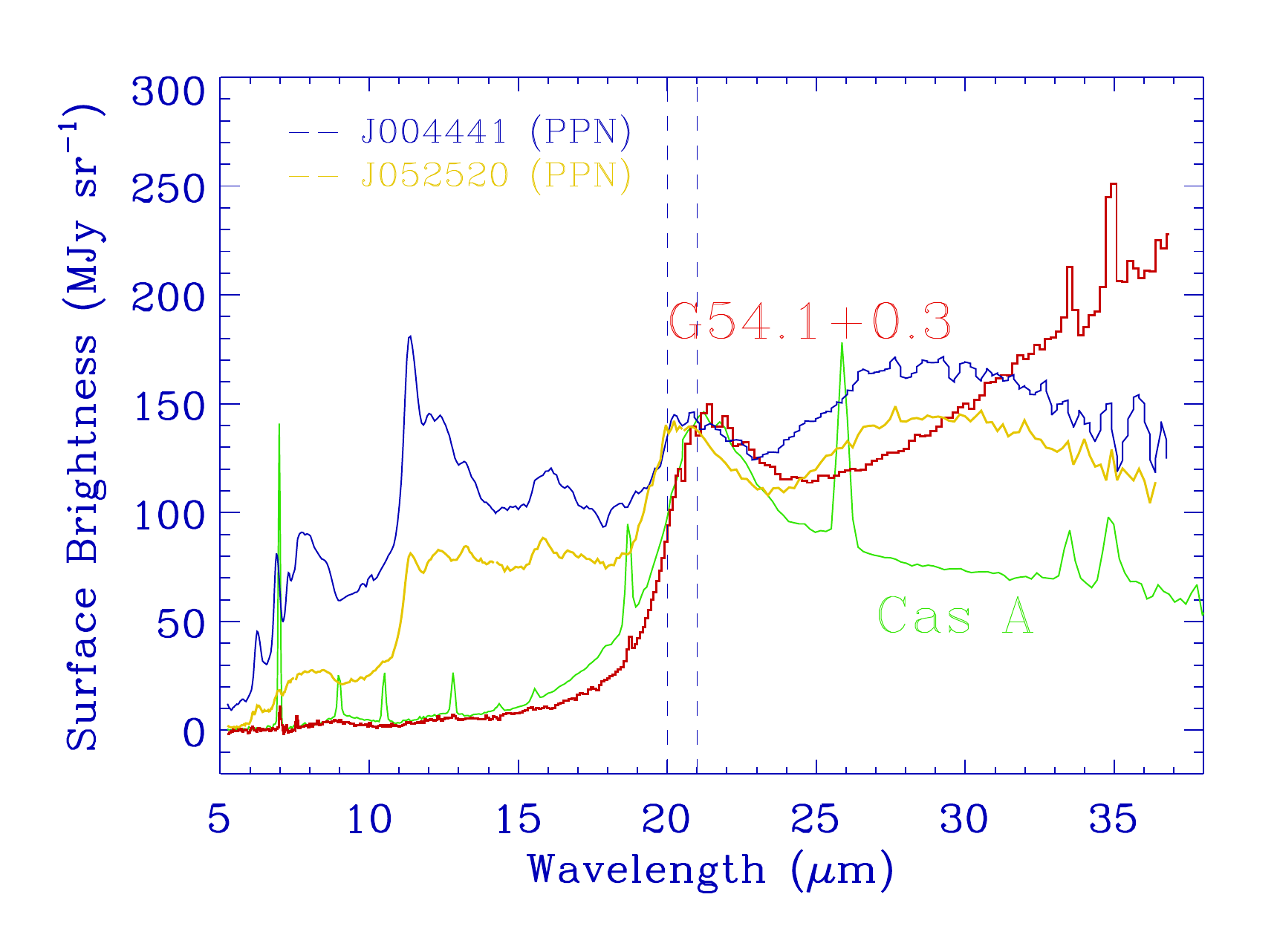}
\caption{Comparing the MIR spectrum for G54.1+0.3 with Cas A and PPNe from
\protect\citet{volk11}. The dotted vertical lines indicate wavelengths
of 20 and 21\mic. Both G54.1+0.3 and Cas A have dust peaks at 21.1\mic\
(the smooth bump) whereas the PPNe (J004441 and J052520) have a dust-peak
feature at 20.1\mic\ as well as broad features at 10-17\mic\ and at
25-35\mic. 
}
\label{fig:g54casairsspec2a}
\end{figure*}

As well as ejecta lines, as discussed in the previous Section, the low-res
spectra show a strong dust feature peaking at 21\mic, with additional
(weaker) dust features at 8.5\mic\ and 10 - 14\mic. In Figure
\ref{fig:g54comparecasa}, we compare the IRS spectrum of Cas A from
\citet{rho08} with G54.1+0.3 and find that the 21\mic\ dust feature is
remarkably similar to that seen in Cas A. (Unlike Cas A, G54.1+0.3's flux
continues to rise well beyond the 21\mic\ peak.) Historically, the term
`21\mic\ dust feature' was allocated to the spectral feature seen in
carbon-rich post-asymptotic giant branch (post-AGB) stars \citep{sloan14},
evolved stars \citep{kwok89} and protoplanetary nebulae (PPNe,
\citealp{volk11,jiang05}); this feature is often accompanied by a peak at
30\mic\ (see also Figure~\ref{fig:g54casairsspec2a}). However, note that the
historical 21\mic\ feature actually peaks between 20-20.3$\mu$m
\citep{kwok89,li13,cerrigone11}. We distinguish this feature (we call
$``$20\mic\ dust feature'') from our 21$\mu$m feature observed in the SNRs
of G54.1+0.3 and Cas A.

\begin{figure*}
\includegraphics{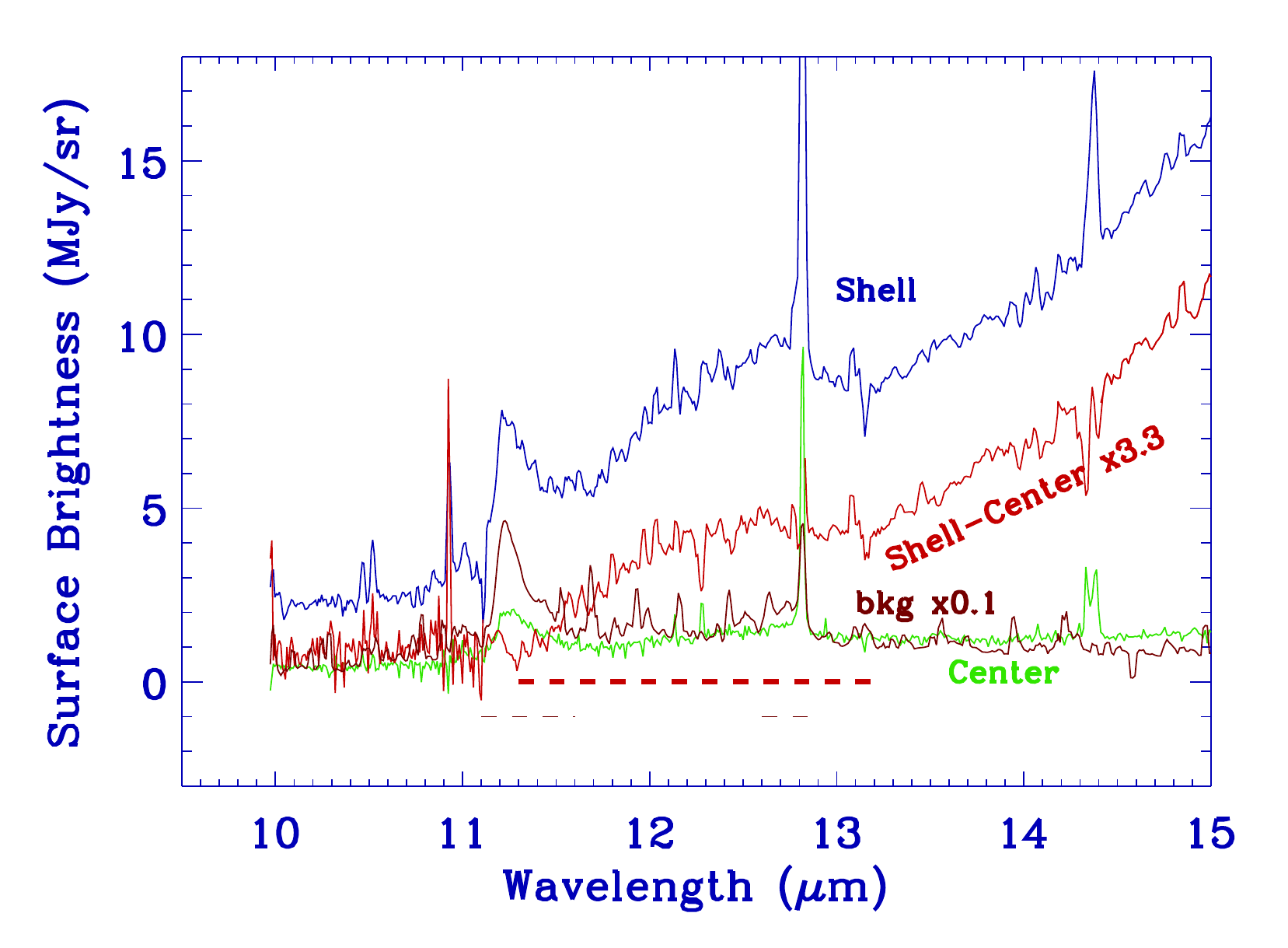}
\caption[]{\spitzer\
IRS spectra of G54.1+0.3. The red curve is the low-res spectrum including
bright emission at the shell. The blue curve is the high-res spectrum from the
shell component of the SNR (the bright `blob' emission in the southwestern
shell), the green is the high-res spectrum measured in the central
region, and the brown is the high-res spectrum from the background (bkg) region. 
The slit locations are shown in Figure \ref{fig:g54IRSslit}. The
line at 14.35$\mu$m shows multiple (high velocity) components in emission
as listed in Table \ref{tab:Tg54irslines}.
The spectra show
dust features at 9, 10-14 (marked in the dotted line in green), and 21$\mu$m
and a narrow dust feature at 11.2$\mu$m.}
\label{fig:g54irsspecwbkg}
\end{figure*}

\begin{figure}
\includegraphics[width=7.5truecm,height=14truecm]{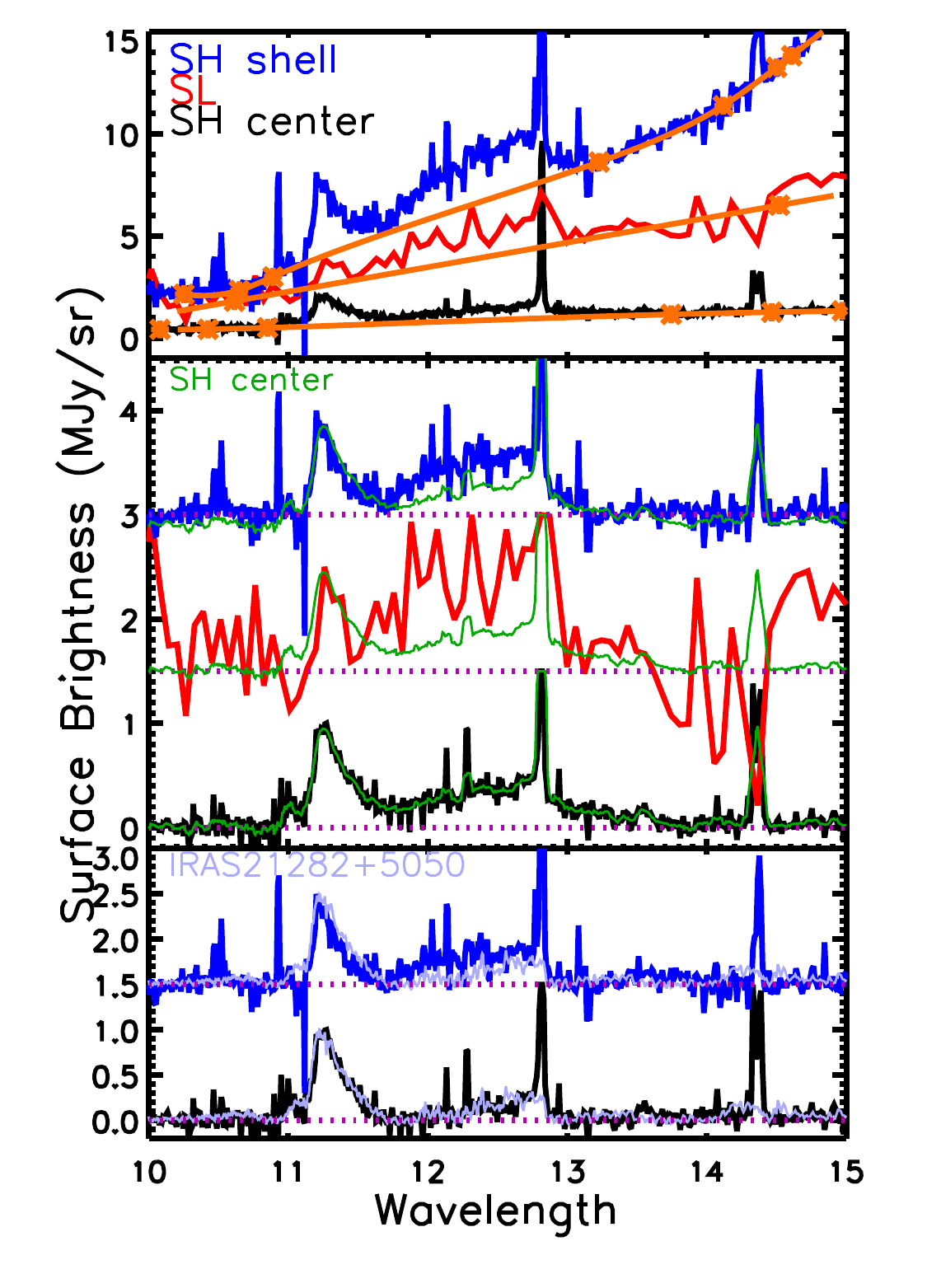}
\caption{
A closer look to the SH and SL spectra of G54.1+0.3 in the 10-14 \mic\ region.
(a: top) The SH and SL spectra with their adopted continua (and chosen
continuum points): a global spline continuum (orange) and a local spline
continuum (light blue; see text for details). (b: middle) A comparison of the
continuum subtracted spectra using the global spline continuum, normalized
to the strength at 11.2 \mic\ (the [NeII] emission line is cut off for
clarity). The spectra are offset from each other by 1.5 units. For
reference, a smoothed version of the continuum subtracted SH center
spectrum is over-plotted in green.  (c: bottom) A comparison of the PAH emission
features in this wavelength range (using a local spline continuum) for the
SH spectra (shell in blue and center in black) of G54.1+0.3 with the spectrum of the 
planetary nebula IRAS21282+5050 \protect\citep[in light blue; taken
from][]{hony01}, normalized to the peak intensity of the 11.2 \mic\
PAH band (the [NeII] emission line is cut off for clarity). The spectra are
offset from each other by 1.5 units.
}
\label{fig:g54em10to14}
\end{figure}

Comparing the \spitzer\ spectra of the two SNRs with PPNe from
\citet{volk11} (Figure~\ref{fig:g54casairsspec2a}), we see that the PPN
feature is offset by $\sim$1\mic\ from the 21.1\mic\ dust feature in
G54.1+0.3  and in Cas A. Unlike the PPNe, neither SNR has a broad dust bump
at 30\mic, though they do have weaker dust features between 10-15\mic\ (see
Section~\ref{sec:11peak}). The 21\mic\ dust feature therefore seems to be
unique to the two young SNRs of Cas A and G54.1+0.3 to date. 
We will return to the dust composition
responsible for this feature in Section~\ref{sec:sedfitting}.

\subsubsection{11$\mu$m dust features }
\label{sec:11peak}

Weaker dust features are seen in the IRS spectra in the 10-14\mic\ range
(Figures~\ref{fig:g54IRSslit} and \ref{fig:g54comparecasa}). In particular,
a broad emission feature is present ranging from approximately 10.5 to
14\mic\ with, on top of it, a strong distinct feature at 11.2 \mic. The
broad feature is much more noticeable in the shell spectrum when we
subtract the center spectrum after scaling the spectra with the sharp
11.2\mic\ feature as shown in Figures~\ref{fig:g54irsspecwbkg} and
\ref{fig:g54em10to14}a. To investigate this emission, we apply a global
spline continuum with anchor points shortward of 11\mic\ and longward of 13
\mic\ (Figure~\ref{fig:g54em10to14}a). The overall shape of the 10-14 \mic\
emission is similar in the two SH spectra while the strength of both the
broad emission feature and the distinct 11.2 \mic\ feature varies
(Figure~\ref{fig:g54em10to14}b).  While less clear in the SL spectrum due
to the lower S/N and spectral resolution, we note that the 10-14 \mic\
emission features are very similar to those in the SH shell spectrum when
allowing for a scaling factor of 0.6 to match the strength of the dust
continuum emission. (The relative strength of the broad emission feature
and the 11.2 \mic\ band is slightly different.) A comparison with the SL
data of Cas A reveals that this broad emission component is also present in
Cas A, albeit at a weaker level, while the distinct feature at 11.2\mic\
seems to be absent (Figure~\ref{fig:g54comparecasa}).  

The distinct feature at 11.2 \mic\ is reminiscent of the
11.2 \mic\ emission band due to Polycyclic Aromatic Hydrocarbon molecules
(PAHs). To explore a possible PAH origin, we follow \citet{hony01} and
apply a local spline continuum: in addition to the anchor points used for
the global spline continuum, two additional anchor points at approximately
11.7 and 13 \mic\ are used (Figure~\ref{fig:g54em10to14}, top panel). We
compare the resulting continuum subtracted SH spectra with the PAH emission
seen in the sample of \citet{hony01}.

The emission seen towards the SH center position is
nearly identical to the PAH emission of the planetary nebula (PN)
IRAS21282+5050: at the position of the 11.0, 11.2 and 12.7\mic\ PAH bands
in the PN, the SH center spectrum shows emission bands with almost
identical shape, e.g., the 11.2$\mu$m PAH feature.
A small discrepancy is also seen at the transition from the 11.0 to the 11.2 \mic\
PAH bands (the 11.0 \mic\ as a shoulder attached to the 11.2 \mic\ band versus two more
distinct bands). Both PAH profiles have been seen in a variety of objects
and within extended sources \citep[e.g.][]{Boersma:13}. The observed
emission towards the SH center position (when subtracting a local spline
continuum) is thus consistent with an origin in PAH emission. Note that the
spectrum at the center was already background subtracted using the
background position 2.5 arcmin away as shown in
Figure~\ref{fig:g54IRSslit}. 

The SH shell spectrum clearly exhibits the 11.2 \mic\ PAH
band and lacks emission near 11.0 and 13.5 \mic. The 11.0 \mic\ is a weaker
PAH band attributed to ionized PAHs \citep{hudgins99} and thus is only
seen for a significant PAH ionization fraction. Hence, the lack thereof
poses no problem to a PAH origin of the 11.2 \mic\ emission. The emission
in roughly 11.8 to 13 \mic\ is somewhat unusual from a PAH perspective.
Clearly, this emission can be consistent with the presence of a 12.7 \mic\
PAH band (at various strengths). However, typically, this feature ranges
from roughly 12.2 to 12.9 \mic\ and a weaker PAH band at 12.0 \mic\ is
discernible (i.e. not blended) when present \citep[see the PN shown in Figure \ref{fig:g54em10to14}
or Figure 7 of][]{hony01}. Note that a typical PAH
emission spectrum ranges between 10 and 15 \mic. 

In contrast, the SH shell spectrum exhibits a smooth increase in surface
brightness starting from $\sim$11.7 \mic\ up to $\sim$13 \mic. A similar
profile is seen towards one or two $\beta$ objects in the sample of
\citet[][see their Fig. 14, e.g. IRAS05360-7121]{matsuura14}.  Therefore,
we conclude that the observed 11.2 \mic\ feature is consistent with an
origin in PAH emission while an origin of (some of) the 11.7-13 \mic\ in
PAH emission is possible, yet uncertain.

While PAH emission is thus present in these SH spectra,
we note that the SH background spectrum exhibits typical and much stronger
PAH emission. This is well illustrated in Figure~\ref{fig:g54irsspecwbkg}.
The 11.0, 11.2 and 12.7 \mum\, PAH emission bands are clearly present and
resemble a typical PAH emission spectrum (see Figure~\ref{fig:g54em10to14}). 
However, this background PAH emission is much stronger than that detected
on-source: the 11.2 \mic\ PAH intensity in the on-source spectra accounts
for 6 and 20\% of that seen in the SH background, respectively, for the
center and shell position. The background observation is taken in staring
mode, resulting in a FOV of 18.4\arcsec x 4.6\arcsec (8$\times$2 pixels).
With a PSF corresponding to 2$\times$2 pixels, we investigate possible
variability along the short slit and only find variation up to the 3\%
level. However, the background region is 2.5 arcmin away from the source.
We therefore searched the Spitzer archive for other IRS observations in the
vicinity to check for possible background variability on a larger scale and
only found one SL observation. Unfortunately, discrepancies between the PAH
band intensities between SL and SH observations \citep[see e.g. Fig. 3 of
][]{Peeters:17} exclude a comparison of the 11.2 \mic\ PAH band intensity
between SH and SL observations at the 20\% level, as required by our
on-source observations. Given the proximity of the center and shell
position to each other (the center positions of the two slits are 25$''$
apart and the slit coverages are next to each other as shown in
Figure~\ref{fig:g54IRSslit}), we assume similar background emission for
both positions. In that case, the maximum residual background emission is
given by the strength of the PAH emission seen in the center position. When
the spectrum of the shell position is corrected for this possible remaining
residual PAH emission, it still exhibits clear PAH emission (in particular
the 11.2 \mic\ band, see Figure~\ref{fig:g54irsspecwbkg}).  

PAH emission gives rise to a number of emission bands, the
major ones centered at 3.3, 6.2, 7.7, 8.6 and 11.2 \mic. We thus expect to
see the 6.2, 7.7 and 8.6 \mic\ PAH bands as well towards the shell.
Unfortunately, these bands are located in the SL
wavelength range where the data are of significantly lower quality and we
do not detect any emission in SL2 (5.5-8.5 \mic) except from the [ArII] and
potential [Ni~I] lines. We compared our SL spectrum with
the IRS spectrum from a well-known PAH source, the reflection nebula
NGC~7023, by normalizing its spectrum so that the integrated strength of the
11.2\mic\ PAH band equals that observed in the SL shell spectrum
(Figure~\ref{fig:g54comparecasa}). Given the strength of the 11.2 \mic\ PAH
band in the SL spectrum, we should have seen PAH emission at 6.2 and 7.7
\mic\ since they are above the noise level. If born out that the 11.2$\mu$m
PAH emission belongs to the SNR, this would be the first detection of PAHs
associated with SN ejecta. This challenges both observations and
theoretical models; none of the molecule and dust formation models
mentioned above have predicted formation of PAHs in SN ejecta.

We extract the profile of the broad emission component by
subtracting the emission seen towards the center from that towards the
shell with the former scaled such that the integrated intensity of the 11.2
\mic\ PAH band is equal to that in the shell. This implicitly assumes the
band profile does not change between the two positions. The resulting
profile is a smooth, symmetric feature from roughly 11.5 to 13 \mic\ shown
in Figure~\ref{fig:g54irsspecwbkg}. A comparison is made between the
center/shell/background + shell-center which is well fitted by a
Gaussian with peak position of 12.3 \mic\ and a FWHM of 1.1 \mic. Possible
carriers of this broad emission feature are PAHs or PAH-related species and
SiC grains. 

The 11.2 and 12.7 \mic\ PAH bands are typically located
on top of a broad emission plateau ranging from roughly 10.5 \mic\ to
roughly 14.5 \mic. While this broad emission plateau shows little variation
towards ISM-type sources \citep[see e.g. Fig. 12 in][]{Peeters:17}, large
variability is seen towards evolved stars exhibiting PAH emission
\citep{sloan14, matsuura14}. Comparing a typical ISM-like PAH plateau with
the broad emission feature seen in the SNR shows that the latter is
considerably more narrow. We explore the possibility that the broad dust
feature between 10-14$\mu$m is from SiC, with spectral fitting described in
Section~\ref{sec:sedfitting} (see  Figures~\ref{fig:G54fitModelA1} and
\ref{fig:casafitModelA1}).

\subsection{Warm and Cool Dust Distribution}

We have generated temperature and dust mass maps using 0.2$^{\circ}$ $\times$ 0.2$^{\circ}$ 
cutouts of the 24$\mu$m, 70$\mu$m, 160$\mu$m, 250$\mu$m, 350$\mu$m, and
500$\mu$m maps. We have re-gridded the maps that are aligned to the same
pixel grid, with 6$''$ pixels, and smoothed each map to the 36$''$
resolution of the 500$\mu$m image, and fit a simple 2-component modified
black-body spectral energy distribution (SED) to each pixel.

\begin{table}
\caption{FIR-submm Flux Densities. $^a$ - The flux is measured within a
circular aperture (with a radius of 90$''$) to cover the entire SNR except
CSO. Since CSO has a limited field of view, we measured the flux from a
circular aperture with the 22$''$ radius (centered on R.A.\ $19^{\rm h}
30^{\rm m} 25.5^{\rm s}$ and Dec.\ $18^\circ$ 52$^{\prime} 13^{\prime
\prime}$, J2000; this is close to the intensity peak of 70$\mu$m image as
shown in Figure \ref{fig:g54submm}.)} \label{tab:Tsubmmflux}
\begin{center}
\begin{tabular}{lllll}
\\
\hline \hline
Data           &  Wavelength & Flux Density \\
               & ($\mu$m)    & (Jy) \\
\hline
\herschel\ PACS  & 70         &91.32$\pm$11.41  \\
\herschel\ PACS  & 100        &70.32$\pm$10.45 \\
\herschel\ PACS  & 160        &29.99$\pm$14.95 \\
\herschel\ SPIRE & 250        &15.08$\pm$3.29 \\
\herschel\ SPIRE & 350        &3.48$\pm$2.07 \\
\herschel\ SPIRE &   500   & 3.32$\pm$1.43 \\
CSO SHARCII$^a$     &  350       &   1.50$\pm$0.30     \\
APEX LABOCA         &   870       &   0.25$\pm$0.04 \\
\hline \hline
\end{tabular}
\end{center}
\end{table}

\begin{figure*}
\includegraphics[width=18truecm]{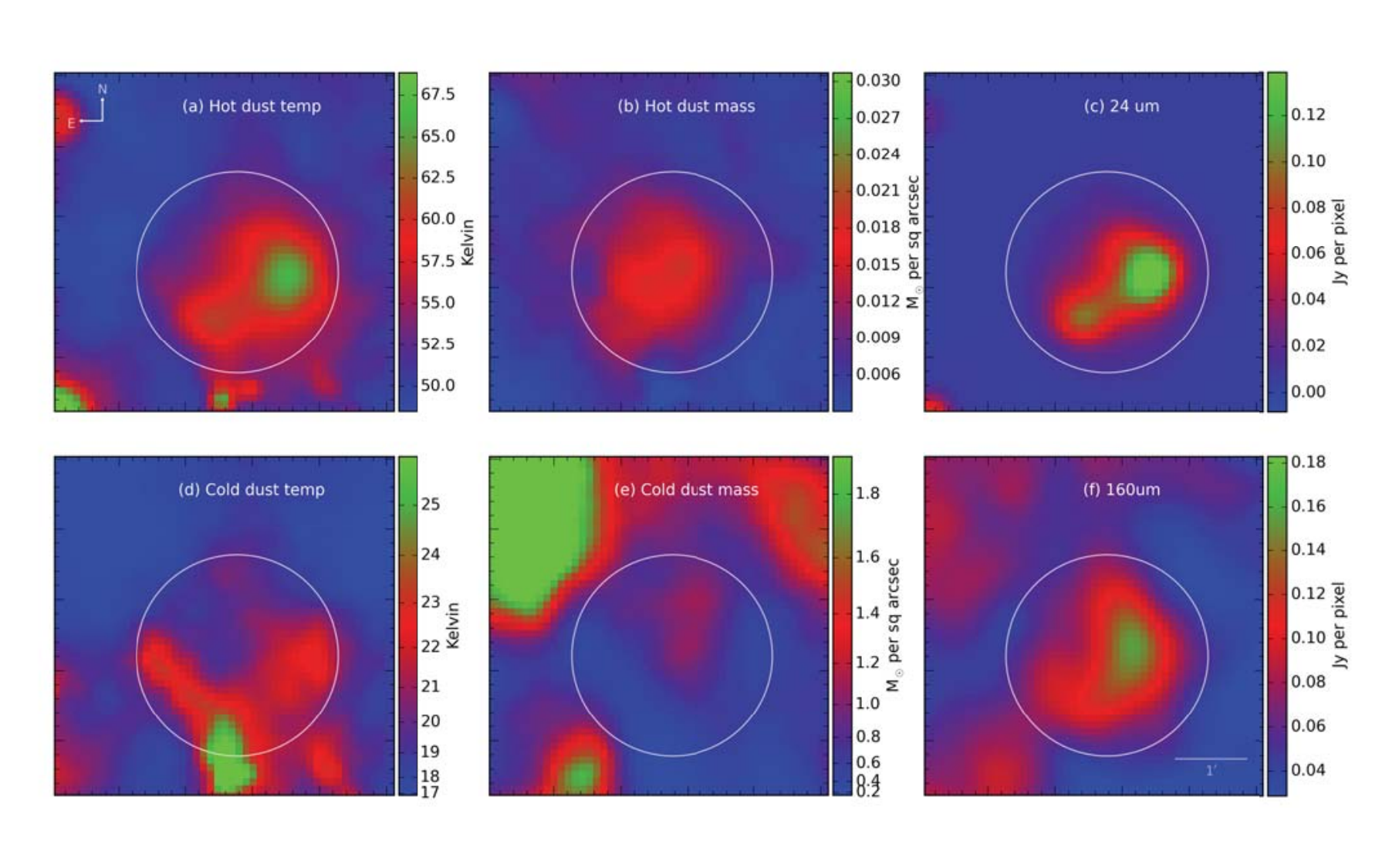}
\caption{
Temperature and dust mass maps of G54.1+0.3 for the hot and cold
components with \spitzer\ 24$\mu$m and Herschel PACS 160$\mu$m maps. (a)
Hot dust temperature map has a range of 50-68\,K, (b) Hot dust mass map has
a range of 0.006-0.02 M$_{\odot}$ per arcsec. (c) \spitzer\ 24$\mu$m image,
(d) Cold dust temperature map has a range of 18-25\,K, and  
 note northeastern shell has a low temperature of 18-20\,K, (e) Cold
dust mass map has a range of 0.4-1.4 M$_{\odot}$ per arcsec and shows cold dust is 
concentrated on the northern part of the SNR and (f)
Herschel PACS 160$\mu$m image.}
\label{fig:g54dustmaps}
\end{figure*}

This method will distinguish between warm and cool dust components.  We
note that a more physical model would account for multiple temperature
components, though in terms of deriving dust masses, the assumption of two
components makes only a factor of ~2 difference in the derivation of dust
mass \citep[see][]{mattsson14}. The fitting was performed using a
chi-squared-minimizing routine which incorporates the color-corrections
for filter response function and beam area (see \citealt{gomez12b} for
details), assumed an emissivity slope of $\beta$=2.0 and used a dust mass
absorption coefficient of $\kappa$ at 500$\mu$m = 0.051 kg m$^{-2}$
\citep{clark2016}. 
Note that for a pixel with an SED that is best fit by a single-component
model (such as pixels with foreground cirrus only), this method will
effectively yield a $`$second' component with no mass (i.e., negligible mass
with a large uncertainty, compatible with zero).

Figure \ref{fig:g54dustmaps} shows temperature and
  dust maps of the hot and cold components. The hot dust mass is
  concentrated toward the central area (slightly to the west) and the
  cold dust mass is concentrated to the northern and central parts of
  the SNR; the maps are consistent with the difference between the
  24$\mu$m and 160$\mu$m images. The temperature where the cold dust
  mass is concentrated ranges from 18 to 25 K as shown in Figure
  \ref{fig:g54dustmaps}d, and the cold temperature extends as low as
  $\sim$18 K in the northeast direction where a thin shell exist in
  the infrared maps (in all bands).  We have attempted to produce dust
  mass and temperature maps with a higher (18$''$) spatial resolution
  using \herschel\ super-resolution maps (e.g. at 500$\mu$m) and
  removed synchrotron emission using a radio map. However, no
  additional information was obtained from that exercise, and because
  the SNR G54.1+0.3 is small, the high-resolution maps may contain
  artefacts. We conclude that the low-resolution maps were sufficient
  for this purpose.

We compare our temperature maps with those by
  \cite{temim17}. We note that \cite{temim17} used observations from
  15$\mu$m to 100$\mu$m for their temperature maps with a resulting
  spatial resolution of 12$''$ and were able to fit only a single dust
  component. In contrast, we used maps from 24 to 500$\mu$m where the
  longer wavelength maps are sensitive to the cold dust, and were able
  to fit two-temperature dust components and generated two temperature
  maps. This results in very different temperatures when comparing the
  two temperature maps. The temperature map by \cite{temim17} ranges
  35-45K range, while our temperature map from the cold component
  spans 18-25K and the map from the warm component spans 50-65K. Their
  map temperatures are compatible with those in our warm temperature
  map given the difference in wavelength coverage.

\section{The Broad-band Spectral Energy Distribution and Dust Features }
\label{sec:sedfitting}

To derive dust masses in G54.1+0.3, we generated a complete
SED from the MIR-submm for G54.1+0.3 (blue data points in
Figure~\ref{fig:G54fitModelA1}) by combining \spitzer\ IRS spectra with
\spitzer, \herschel, SHARC-II and LABOCA photometry. The low-resolution
\spitzer\ spectra was used as it provides extra coverage over the 5-8\mic\
range and highlights the broad dust features in the NIR and MIR. To combine
the spectra with the broadband SED, we determined the total flux density
(in Jy) of the entire SNR using \spitzer\ 24\mic\ image and scaled the low-res
IRS spectrum to match the \spitzer\ 24\mic\ flux.  
There is no significant change in the spectrum along the slit, and 
the IRS spectrum at relatively short wavelengths is to constrain the dust compositions
and the dust mass is largely constraint by far-IR data points from Herschel photometry.  
The estimated flux densities of G54.1+0.3
in submm and far-infrared are summarized in Table
\ref{tab:Tsubmmflux}\footnote{The flux density of the SNR at 350$\mu$m in
the SHARC-II and \herschel\ SPIRE images are comparable within the flux
errors.}.  At the long IR wavelengths, the largest contribution to the flux
uncertainty can originate from the background variation. We varied the
choice of background region when measuring the flux density of G54.1+0.3 at
various wavelengths and found the flux at 350\mic\ and 500\mic\ changed up
to 20\% depending on the location of the background aperture. This is due
to increasing amount of interstellar cirrus in the vicinity of the SNR at
the longest wavelengths. The FIR-submm fluxes along with their
uncertainties are listed in Table~\ref{tab:Tsubmmflux}.

As G54.1+0.3 is Crab-like, radio-bright SNR, synchrotron emission could be
an important contributor at FIR-submm wavelengths particularly at
500-870\mic\ (as seen in the Crab, \citealp{gomez12b}).  Using the Very
Large Array (VLA) map at 1\,GHz with spectral slope $S_{\nu} \propto
\nu^{-0.16}$ \citep{leahy08, velusamy88, lang10} and integrating the flux
within the same aperture used to measure the submm flux, we estimate that
$\sim$0.15\,Jy at 870\mic\ is likely to originate from synchrotron
($\sim$60\% of the total flux). At 350\mic, we would expect 0.13\,Jy of
flux due to synchrotron emission ($<$10\%). However, comparing the
structure of the submm emission seen in the 350 and 870\mic\ bands
(Figure~\ref{fig:g54submm}), there is very little similarity between the
distribution of the emission seen in radio and the 24\mic\ or submm
Figures~\ref{fig:g54multi} and \ref{fig:g54submm}). Therefore, we
excluded 870$\mu$m data in the spectral fitting, but overplotted in the fit
results in Figures \ref{fig:G54fitModelA1}, \ref{fig:g54modelfitsa} and
\ref{fig:g54modelfitsb}. Hereafter, we fit the dust SED after freezing a
synchrotron model with the spectral index of 0.16 and a flux density of
0.364 Jy at 1 GHz. The flux at 870$\mu$m is slightly  above the synchrotron
radiation or is consistent with the radiation within the error.

\begin{figure}
\includegraphics[scale=1.2,angle=0,width=8.9truecm,height=6.4truecm]{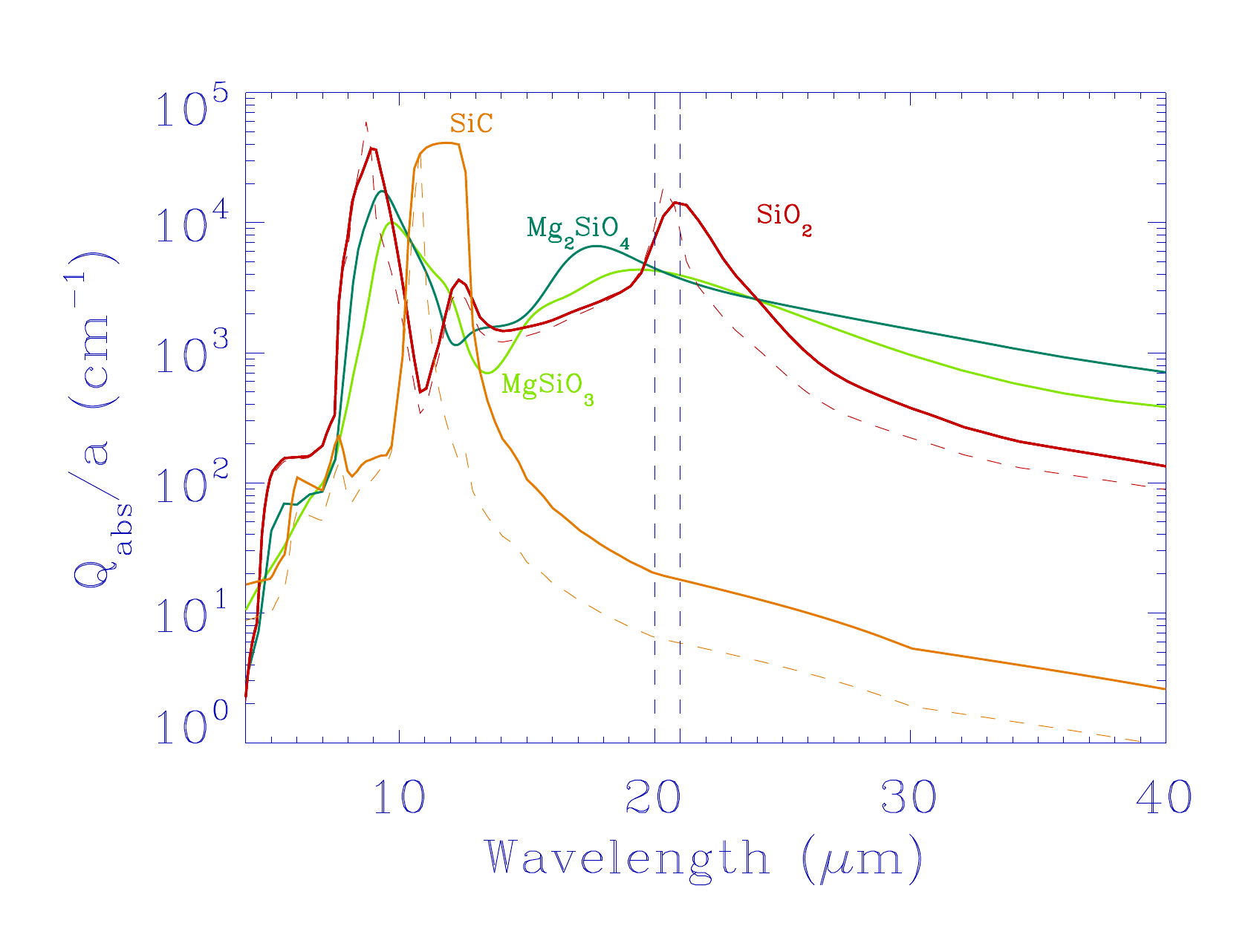}
\caption[]{Optical constants for spherical silicate dust
grains relevant for this work, including spherical grains of SiO$_2$ (red
dotted line), SiC (orange dotted line), MgSiO$_3$ (green), and
Mg$_2$SiO$_4$ (dark green). The continuous distribution of ellipsoidal
(CDE) models for SiO$_2$ (red solid line) and SiC (orange solid line) are
also shown (see the text for details). }
\label{fig:Qabssio2etc}
\end{figure}

\begin{figure*}
\includegraphics[width=14.5truecm]{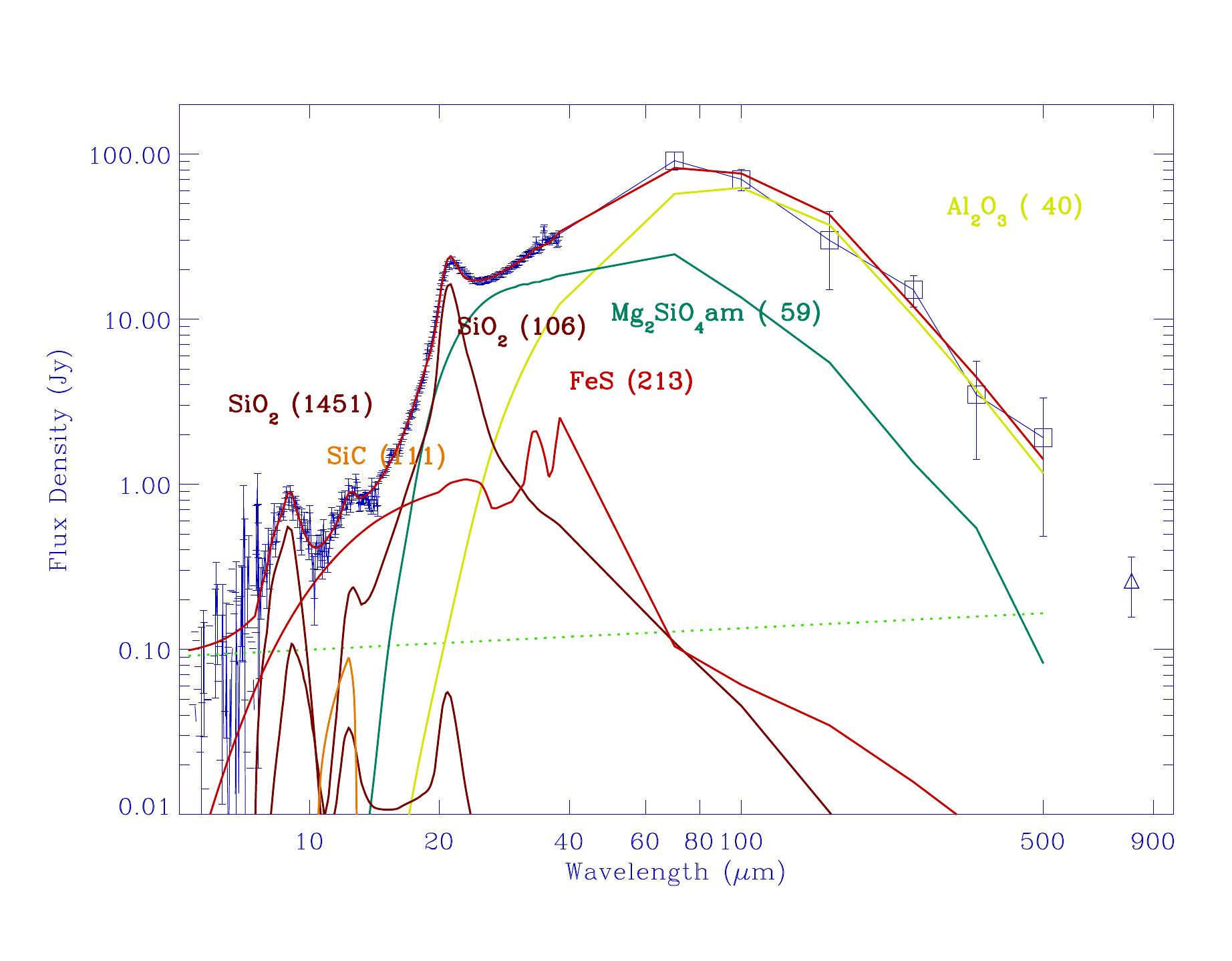}
\caption{The spectral energy distribution of the SNR G54.1+0.3 from MIR -
submm wavelengths (the broadband photometry from Table~\ref{tab:Tsubmmflux}
and the IRS spectra are shown by blue data points) superposed on the dust
fit of model A1 (see Table ~\ref{tab:Tdustmass}). The data and the total
fit are shown in blue and thick red lines, respectively, and \herschel\
fluxes are marked with squares. The dust temperatures are shown in
parentheses, and the dotted lines are from the second temperature
components. Synchrotron continuum contribution (green dashed line) is
estimated based on the radio fluxes . The 21$\mu$m and 11$\mu$m dust
features are  well produced by the CDE grain model of Silica and Silicon
carbide (SiC), respectively. As well as the dust features and compositions
required to fit the IRS spectrum of the SNR, we require a population of
colder dust grains to explain the fluxes at wavelengths beyond 70\,$\mu$m
(see Section~\ref{sec:sedfitting} for details). The different grain
compositions required and fitting properties highlighted individually in
this plot are listed in full in Table~\ref{tab:Tdustmass}.
}
\label{fig:G54fitModelA1}
\end{figure*}

In an attempt to obtain a better fit to the SED of G54.1+0.3, we use the
same SED fitting routine used for Cas A, but now try different grain
compositions and different grain models.  Specifically, we use the
continuous distributions of ellipsoidals (CDE) grain model for SiO$_2$
(silica) and SiC (Silicon carbide) \citep{bohren83}; the absorption
efficiency for this grain species and others is shown in
Figure~\ref{fig:Qabssio2etc}. The first results of fitting using SiO$_2$ to
reproduce the 21$\mu$m dust feature in Cas A were presented in conference
proceeding \citep{rho09proc}. The dust continuum is fit with the Planck
function [B$_{\nu}(T)$] multiplied by the absorption efficiency ($Q_{abs}$)
for various dust compositions, varying the amplitude and temperature of
each component. The  optical constants of the grain species used in the
calculation are the same as those of  \cite{hirashita05}, except for
amorphous SiO$_2$ \citep{philipp85}, amorphous Al$_2$O$_3$
\citep{begemann97}  and we apply Mie theory \citep{bohren83} to calculate
the absorption efficiencies, Q$_{abs}$, assuming the grains are small
a$\leq$0.01 $\mu$m \citep{nozawa13, nozawa10}.

The dust mass of $i$-grain type is given by: $$ M_{dust,i} = {F_\nu^i \,
d^2 \over {B_\nu (T_{d,i}) \, \kappa_i } }= {F_\nu^i d^2 \over
{B_\nu(T_{d,i})}} { 4 \, \rho_i \, a \over {3 \,Q_{abs, i}}} $$ where
$F_\nu^i$ is the flux from $i$-grain species, $d$ is the distance,  $B_\nu$
is the Planck function, $\rho_i$ is the bulk density, and $a$ is the dust
particle size. We fit the flux density for each spectral type using scale
factors $C_i$ for each grain type $i$, such that F$_{\nu}^i$ = $\Sigma_i \,
C_i \, B_\nu \, Q_{abs,i} / a$, where Q($\lambda$, a) are calculated using
Mie theory \cite{bohren83} and assuming a relatively small grain size a =
0.01 $\mu$m \citep{nozawa13}.
Note that
the calculated values  of $Q_{abs} /a$ are independent of the grain size
(a) for a $<<$0.1 $\mu$m at these wavelengths (i.e. as long as 2$\pi |m|
a/\lambda$ $<<$1 where $m$ is the complex refractive index). Thus the
derived scale factor C$_i$ as well as the estimated dust mass (see
Section~4.4) are independent  of the radius of the dust. We applied this
technique in our previous papers by \cite{rho09, arendt14, rho08}.  The
dust compositions of the best fits are summarized in Table
\ref{tab:Tdustmass}.

We have quantified the goodness of fit ($\chi^2$) for the models using the
{\sc mpfit} IDL routine \citep{markwardt09}, though we note that there are
degeneracies between the many different grain species that can fit the SED.
 The different models we tried with different grain compositions are
described in full in Table~\ref{tab:Tdustmass}. Note that the goodness of fit
may depend on initial selection of grain composition.

\begin{figure}
\includegraphics[width=8.5truecm]{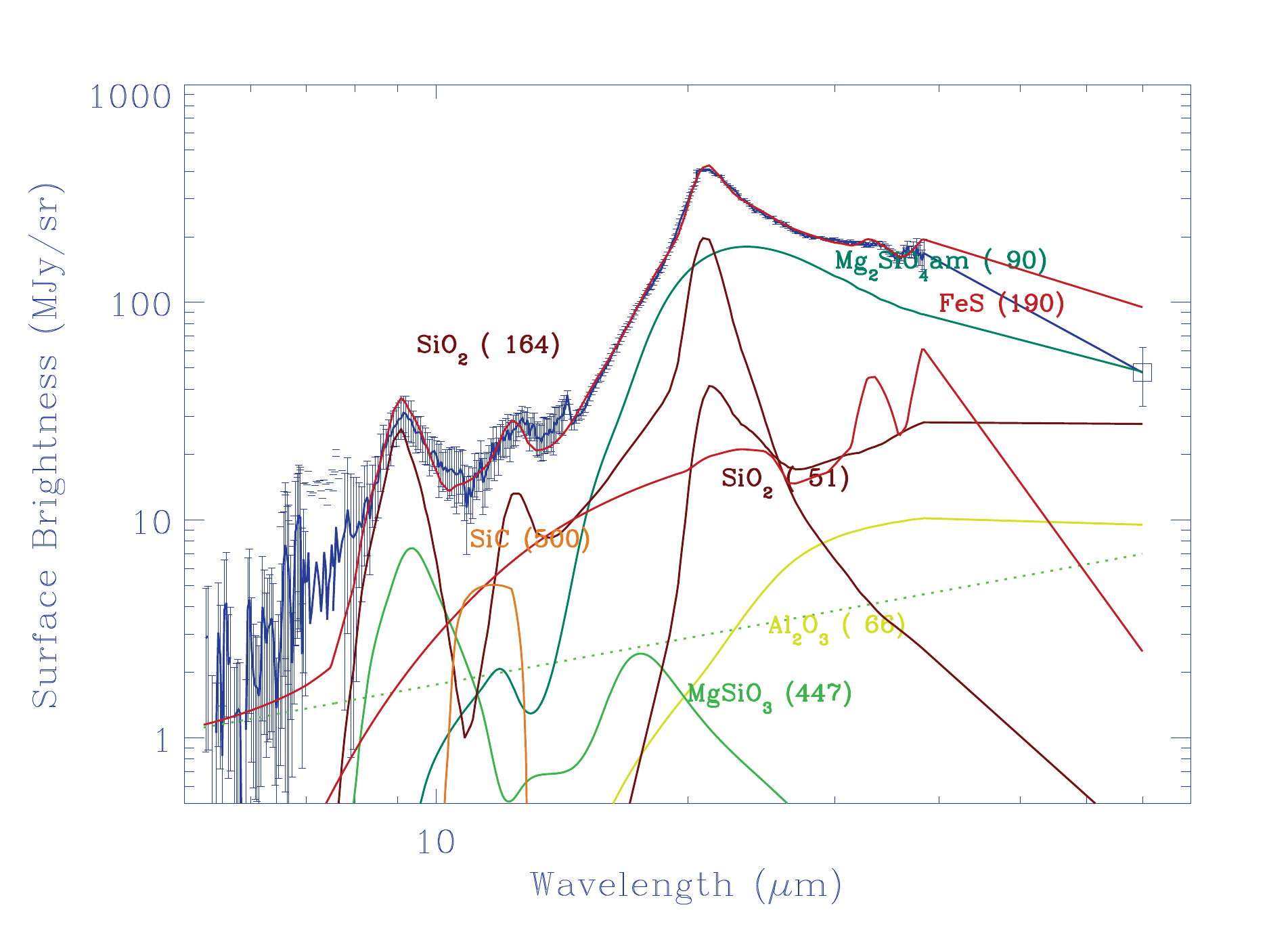}
\caption{Cas A spectral fitting using the Model A1 (Table~\ref{tab:Tdustmass}), the
same model as that of G54.1+0.3 to the \spitzer\ SED of Cas A from
\citet{rho08}. The model includes using the continuous distribution of
ellipsoidal models instead of spherical grains for SiO$_2$ and SiC, which
fairly well reproduce the 21$\mu$m and 11$\mu$m dust features, respectively.
}
\label{fig:casafitModelA1}
\end{figure}

\subsection{Silica for 21$\mu$m dust and Silicon Carbide for 11$\mu$m dust features}

\cite{rho08} fitted the IRS spectra and
21\mic\ dust peak feature using a combination of grains that included
SiO$_2$ (silica), Mg protosilicates, and FeO grains (their so-called Model
A). However, \citet{rho08} noted that the 21\mic\ feature was not perfectly
reproduced with this model as shown in Figure 3 of their paper.
Here we attempt to do a similar
analysis for G54.1+0.3. We first performed spectral fitting of the
21$\mu$m-peak dust using the CDE grain model of SiO$_2$ and spherical model for
MgSiO$_3$. The comparison of absorption coefficient between the spherical
and CDE grain model in Figure \ref{fig:Qabssio2etc} show that the SiO$_2$ and SiC
CDE features are smoother, broader feature than those of spherical grain
models.  Here we find that the 21 $\mu$m peak dust for spectra of G54.1+0.3
is reasonably well fit by SiO$_2$ in Figure \ref{fig:G54fitModelA1} without
requiring Mg proto-silicates nor FeO in contrast to the original Cas A
study. In addition, the newly identified 11-12.5\mic\ feature seen here
(Figure~\ref{fig:g54comparecasa}) can be fit with the SiC CDE grain model.

For a SiC grain of a given size "a" and a given 21$\mu$m feature strength ($Q_{abs}$/$a$),
\citet{jiang05} calculate its equilibrium temperature
and its emission spectrum to generate both dust feature at 11.3 and 21$\mu$m.
However, the required strength for the 21$\mu$m feature is
stronger than that for the 11.3 $\mu$m feature of SiC.
Because the low-res spectrum covered an additional wavelength coverage at 5-10
$\mu$m which include the 9$\mu$m dust feature
(see Figure~\ref{fig:g54casairsspec2a}), we used the low-res
IRS spectrum together with Herschel photometry data for dust spectral
fittings.

After we fit the CDE models of SiO$_2$ and SiC to reproduce the 21$\mu$m
and 11$\mu$m dust features obtained in the low-res spectra of G54.1+0.3, we
incorporate other grain compositions based on Model A from \citet{rho08}
including amorphous FeS (150\,K) and Mg$_2$SiO$_4$ (58\,K) to account for
the emission between 5-30$\mu$m, and Al$_2$O$_3$ to reproduce the spectra
beween 30-500$\mu$m\footnote{The LABOCA data point at 870$\mu$m is
effectively treated as an upper limit because a portion of the 870$\mu$m
may be from foreground material of molecular clouds.}. We also include
grains of SiO$_2$ at high temperature (300\,K) to account for the emission
feature around 9.8\mic, but this component is added to mimic IR
emission from stochastic heating (e.g. \citealt{rho08, rho09, lagage96}). This model
is shown in Figure~\ref{fig:G54fitModelA1} (Model A1); the fit produces a
reduced $\chi^2$ of 3.7 and yields a total dust mass for the SNR of 0.16
M$_\odot$ (Table~\ref{tab:Tdustmass}).

 \begin{figure}
 \centering
\includegraphics[width=8.5truecm]{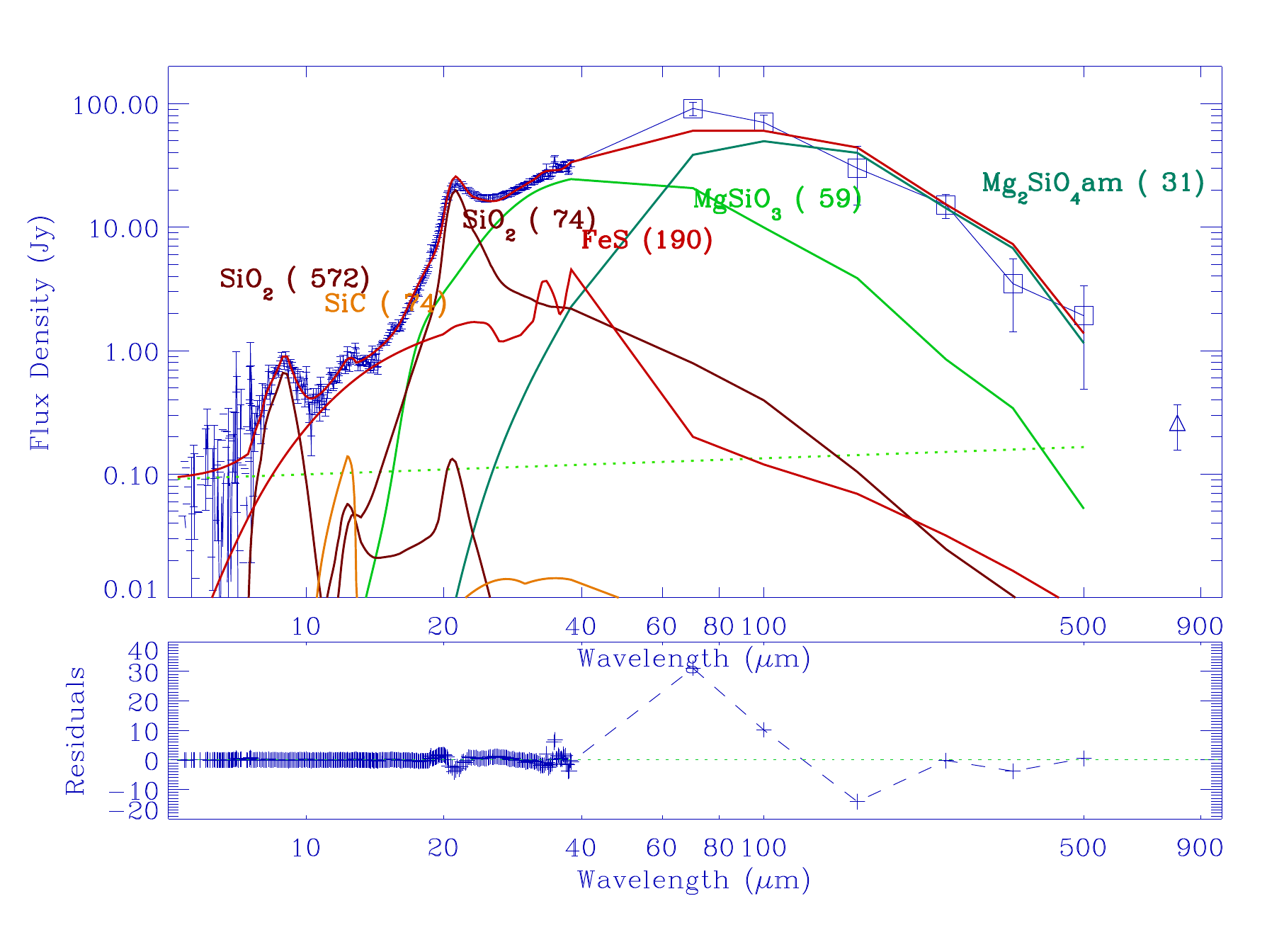}
\includegraphics[width=8.5truecm]{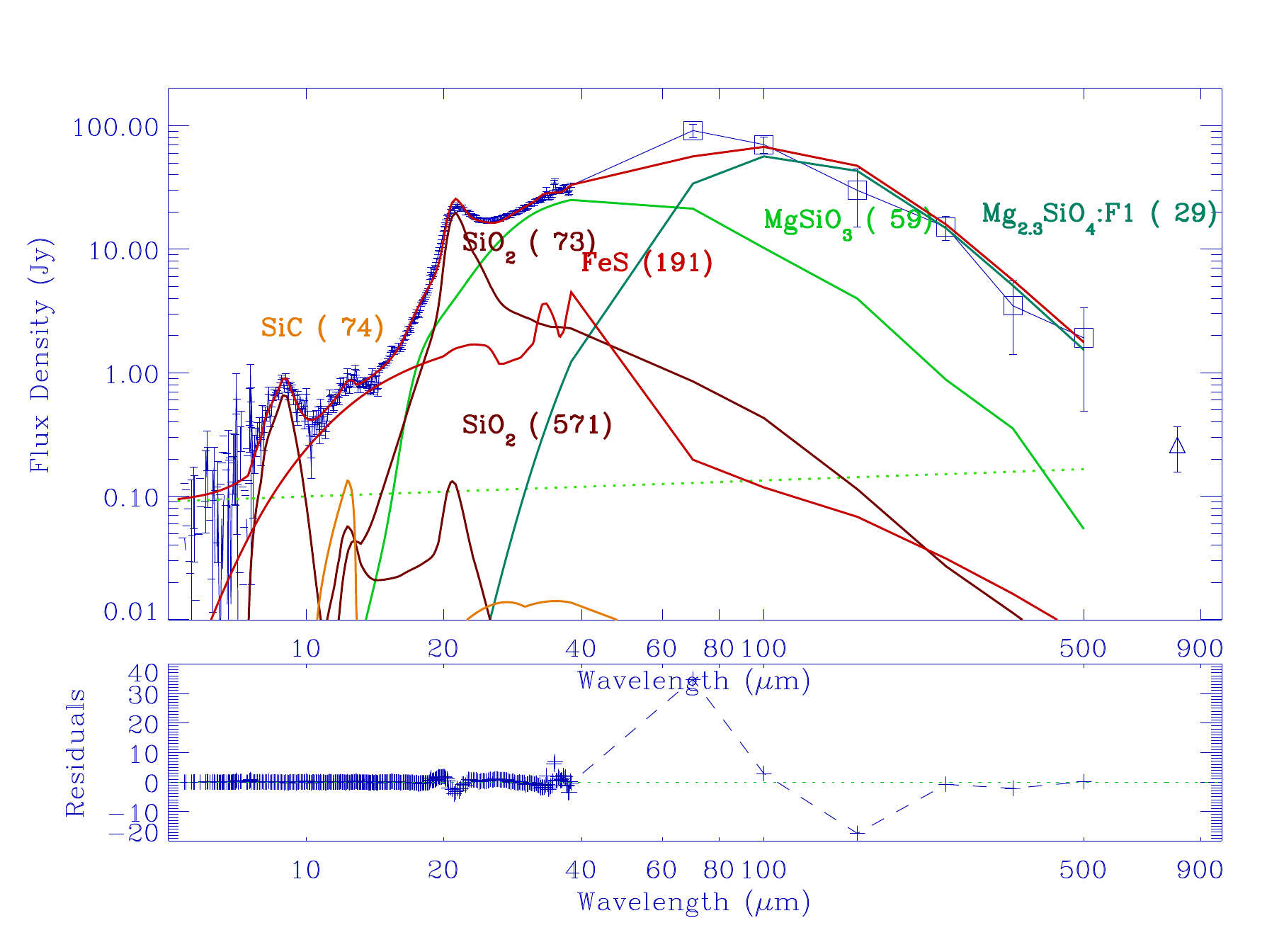}
\caption{Spectral energy distribution of G54.1+0.3 superposed on the dust fit of a
warm dust component made of MgSiO$_3$, with cold dust of Mg$_2$SiO$_4$
({\it Top:} Model A2) and of Mg$_{2.3}$SiO$_4$ using its Q$_{abs}$ at 30 K
({\it Bottom:} model A2F1), respectively. Silicate dust using a
temperature-dependent (30\,K) Q$_{abs}$ produced a smaller dust mass (see
Table \ref{tab:Tdustmass}).   
}
\label{fig:g54modelfitsa}
\end{figure}

\begin{figure}
\includegraphics[scale=1.2,angle=0,width=8.9truecm,height=6.4truecm]{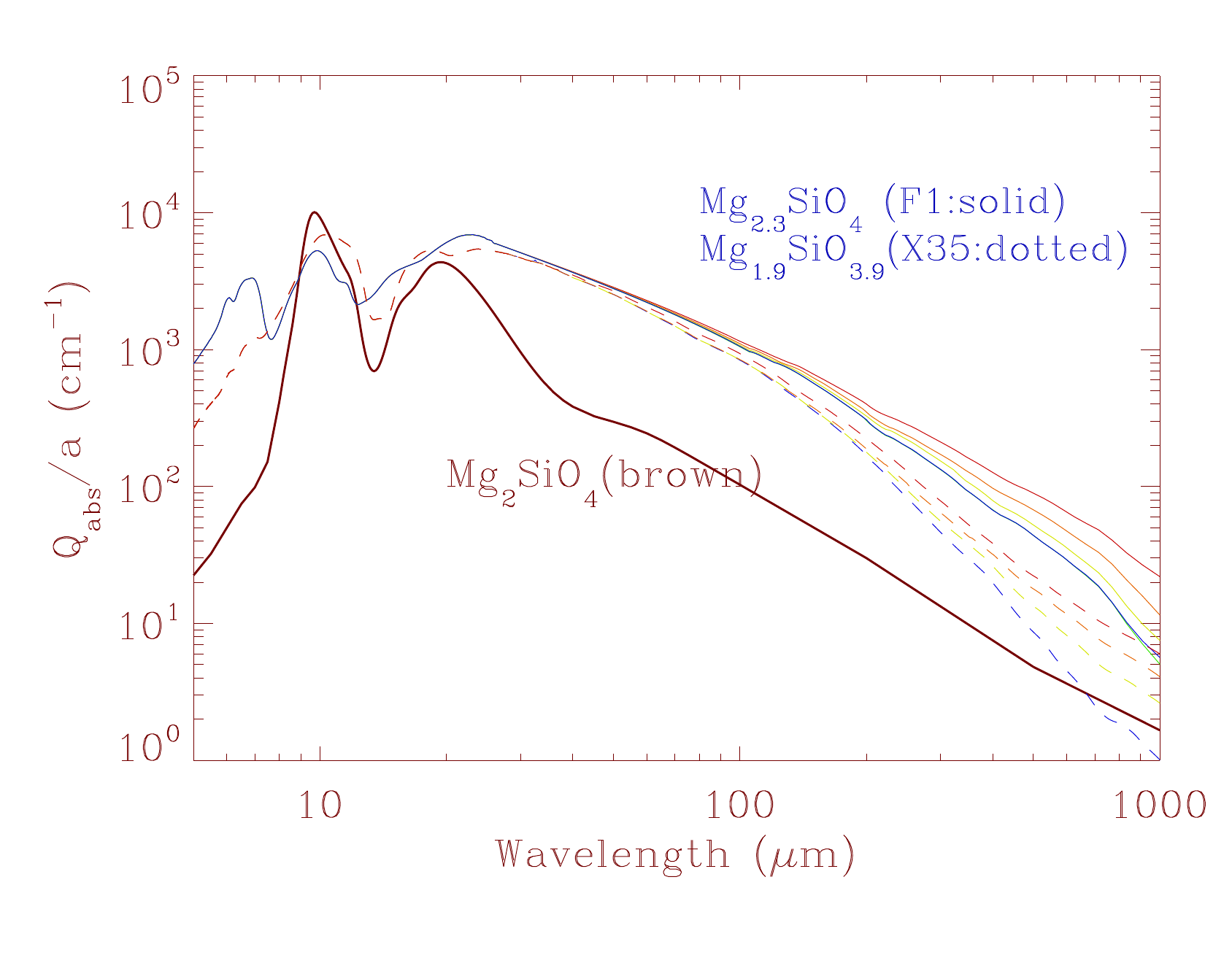}
\caption{
Optical constants for temperature dependent CDE models Mg$_{2.3}$SiO$_4$
(F1) and Mg$_{1.9}$SiO$_{3.9}$ (X35) (Demyk et al. 2013; Coupeaud et al.
2011) are compared with Mg$_2$SiO$_4$. For each sample of F1 and X35, the
spectra are measured for a grain temperature of 300 K (red), 200 K
(orange), 100 K (light green) and 30 K (blue).}
\label{fig:Qabsmg2sio4Tempe}
\end{figure}

As Model A1 is a good fit to G54.1+0.3, we next determine if this updated
combination of grain compositions can also provide a better fit to the Cas
A spectra compared to \citet{rho08}. Figure~\ref{fig:casafitModelA1} shows
that indeed the CDE SiO$_2$ and SiC grains provide a better fit for the
21$\mu$m and 11$\mu$m dust features of Cas A, removing the residual seen in
the IR in the earlier study. We note however that the IRS spectral coverage
for G54.1+0.3 is limited to the center and south-western shell, and full
spectral coverage may reveal different nucleosynthetic layers like those in
Cas A.
\cite{temim17} fit the 21$\mu$m dust feature of G54.1+0.3 with
Mg$_{0.7}$SiO$_{2.7}$ and stated that they could not fit the 21$\mu$m
feature using SiO$_2$ grains. We note that the shape of absorption
coefficient of SiO$_2$ \citep[Figure 4 of][]{temim17} is different from the
one we used in Figure \ref{fig:Qabssio2etc}. Our fits with SiO$_2$
reproduced the 21$\mu$m dust feature very well and the theoretical models
\citep{nozawa03, bianchi07, cherchneff10} predict production of silica in
SN ejecta, and laboratory measurement \citep{haenecour13} discovered two
SiO$_2$ grains that are characterized by and originated from Type II SN
ejecta.

\begin{table*}
  \footnotesize{
\caption[]{Dust Spectral fitting Results and Estimated Dust Mass. $^a$ - The 11$\mu$m dust feature includes a contribution from 
both SiC and hot dust composed of SiO$_2$. $^b$ - The three best solutions are in bold. $c$ - $\Delta$$\chi^2$ in the parenthesis 
is the reduced $\chi^2$ using statistical errors as weights. $^d$ - The number in the parenthesis is the dust temperature yielded 
from the spectral fitting for G54.1+0.3. Temperatures greater than
$\sim$1800\,K is from stocastic heating of the grains. 
}\label{tab:Tdustmass}
\begin{center}
\begin{tabular}{llrrrrrl}
\hline\hline
\multicolumn{1}{l}{Model$^a$} & \multicolumn{1}{l}{$\Delta \chi^2$} & \multicolumn{2}{c}{Dust Features} & \multicolumn{3}{c}{Dust components} & \multicolumn{1}{c}{Dust} \\
\multicolumn{1}{l}{} & \multicolumn{1}{l}{($\chi^2$/dof)} &\multicolumn{1}{r}{21$\mu$m } &\multicolumn{1}{l}{11$\mu$m$^a$} &
\multicolumn{1}{c}{Warm dust I} & \multicolumn{1}{l}{II [mass]}& \multicolumn{1}{c}{Cool dust} &  \multicolumn{1}{c}{mass} \\
\multicolumn{1}{c}{} & \multicolumn{1}{c}{} &\multicolumn{1}{r}{(K)} &\multicolumn{1}{c}{(K)} &
\multicolumn{1}{r}{ (K)} &\multicolumn{1}{l}{II (K) [M$_\odot$]} & \multicolumn{1}{l}{(K) [M$_\odot$]} & \multicolumn{1}{r}{ (M$_\odot$)} \\ 
\hline
{\bf A1$^b$} & {\bf 3.70 (1.44$^c$)}  & {\bf SiO$_2$ (106, 1450$^d$)}   &  {\bf SiC (110)}  & {\bf FeS (213)} &{\bf Mg$_2$SiO$_4$ (59) [0.026]}  & {\bf Al$_2$O$_3$ (40) [0.13]}   &{\bf 0.16 } \\
               &                      &   ($\pm$5, $\pm$1400)          & ($\pm$5)          &  ($\pm$5)     &  ($\pm$1) [$\pm$0.004]   & ($\pm$2) [$\pm$ 0.03]  & ($\pm$0.04)\\
{\bf A2} & {\bf 3.64 (4.42)}  & {\bf SiO$_2$ (75, 572)$^d$} & {\bf SiC (74)} &{\bf FeS (190)} &{\bf MgSiO$_3$ (59)} [0.019] & {\bf Mg$_{2}$SiO$_{4}$ (31)} [0.879] &  {\bf 0.90 } \\
               &                      &   ($\pm$2, $\pm$5)     & ($\pm$2)    &  ($\pm$3)      & ($\pm$1) [$\pm$0.003] &($\pm$2) [$\pm$0.250]   &  [$\pm$0.30] \\
A2F1 & 4.60 (5.10)  & SiO$_2$ (73, 571)) [0.0019] & SiC (74) &FeS (191) & MgSiO$_3$ (59) [0.020] & Mg$_{2.3}$SiO$_{4}$(29) [0.147]   &0.17  \\
     &              &   ($\pm$2, $\pm$5) [$\pm$0.0003]    & ($\pm$1)  &  ($\pm$3)&  ($\pm$1) [$\pm$0.002]   & ($\pm$2) [$\pm$0.036] & ($\pm$0.04) \\
A2X35 & 4.60 (5.32) & SiO$_2$ (74, 571)) [0.0021] & SiC (74) &FeS (191) & MgSiO$_3$ (59) [0.020] & Mg$_{1.9}$SiO$_{3.9}$(27) [0.334]   &0.36  \\
    &               &   ($\pm$2, $\pm$5) [$\pm$0.0003] & ($\pm$2)          &  ($\pm$3)     &  ($\pm$1) [$\pm$0.003]   & ($\pm$2) [$\pm$0.102] &($\pm$0.11) \\
A3 & 4.89 (6.81) & SiO$_2$ (69, 200) [0.003] & SiC (141)  &FeS (189)  & MgSiO$_3$ (59) [0.019] & Al$_2$O$_3$ (30) [0.326]   & 0.35  \\
        & &   ($\pm$2, $\pm$11) [$\pm$0.0008] & ($\pm$88 &  ($\pm$4)  &  ($\pm$1) [$\pm$0.0015] & ($\pm$2) [$\pm$0.08] & ($\pm$0.09) \\
{\bf A4} & {\bf 3.72 (1.06)} &  {\bf SiO$_2$ (109, 2793)} [8E-5] & {\bf SiC (137)} & {\bf FeS (213)} & {\bf Mg$_2$SiO$_4$ (58) [0.033]} & {\bf carbon (39) [0.223]}  & {\bf 0.26} \\
        &                   &   ($\pm$3, $\pm$92) [$\pm$2E-5]    & ($\pm$84)     &  ($\pm$4)  &  ($\pm$2) [$\pm$0.004]& ($\pm$1) [$\pm$0.049] & ($\pm$0.05)\\
A5 & 3.96 (1.69) & SiO$_2$ (110, 2621) [8E-5]  & SiC (276)  &FeS (203)  & Mg$_2$SiO$_4$ (55) [0.054] & Fe$_3$O$_4$ (42) [0.022]   & 0.08  \\
   & &   ($\pm$2, $\pm$89) [$\pm$1E-5] & ($\pm$182) &  ($\pm$4)     &  ($\pm$1) [$\pm$0.007]   & ($\pm$2) [$\pm$0.004] &($\pm$0.01) \\
\hline
{\bf B1}&{\bf 3.57 (1.30) }& {\bf SiO$_2$ (113, 2686) [9.4E-5] }& {\bf SiC (102) } & {\bf C (180)} &{\bf Mg$_2$SiO$_4$ (62) [0.018]} & 
{\bf Al$_2$O$_3$ (42) [0.118] }&{\bf 0.14 } \\
  &            &   ($\pm$1, $\pm$1326) [2.8E-5]  & ($\pm$48)    &  ($\pm$4) &  ($\pm$1) [$\pm$0.003]  & ($\pm$1) [$\pm$0.008] &($\pm$0.01) \\
B2 & 4.56 (3.62)  & SiO$_2$ (75, 557) [1.57E-3] & SiC (74) [8.0E-4]& C (156) [1.2E-5]  & MgSiO$_3$ (56) [0.021] & Mg$_2$SiO$_4$ (32) [0.830]   &0.85  \\
        &  &   ($\pm$2, $\pm$4) [0.26E-3] &($\pm$1) [$\pm$0.5E-4]&($\pm$2) [0.1E-5]& ($\pm$1) [$\pm$2.5E-3]   & ($\pm$2) [$\pm$0.195] & ($\pm$0.20) \\
B3 & 4.36 (4.76)  & SiO$_2$ (76, 509) [1.45E-3] & SiC (65) [7.9E-3]  &C (160) & MgSiO$_3$ (59) [0.021] & Al$_2$O$_3$ (33) [0.276]   & 0.31  \\
   &      & ($\pm$2, $\pm$75) [0.25E-3]    & ($\pm$1) [$\pm$0.3E-3]   &  ($\pm$2)  &  ($\pm$1) [$\pm$0.002]   & ($\pm$2) [$\pm$0.061]  & ($\pm$0.06)\\
B4 & 4.18 (3.17) & SiO$_2$ (113, 2719) [7E-5] & SiC (483)  &C (169)  & Mg$_2$SiO$_4$ (54) [0.073] & Fe$_3$O$_4$ (44) [0.0089] & 0.08 \\
   &      &   ($\pm$2, $\pm$87) [$\pm$1E-5]  & ($\pm$25)       &  ($\pm$5)     &  ($\pm$1) [$\pm$0.008]   & ($\pm$3) [$\pm$0.001] & ($\pm$0.01) \\
\hline \hline
\end{tabular}
\end{center}
}
\renewcommand{\baselinestretch}{0.8}
\end{table*}

\subsection{Exploring other SED fitting models}

\begin{figure*}
\centering
\includegraphics[width=13truecm]{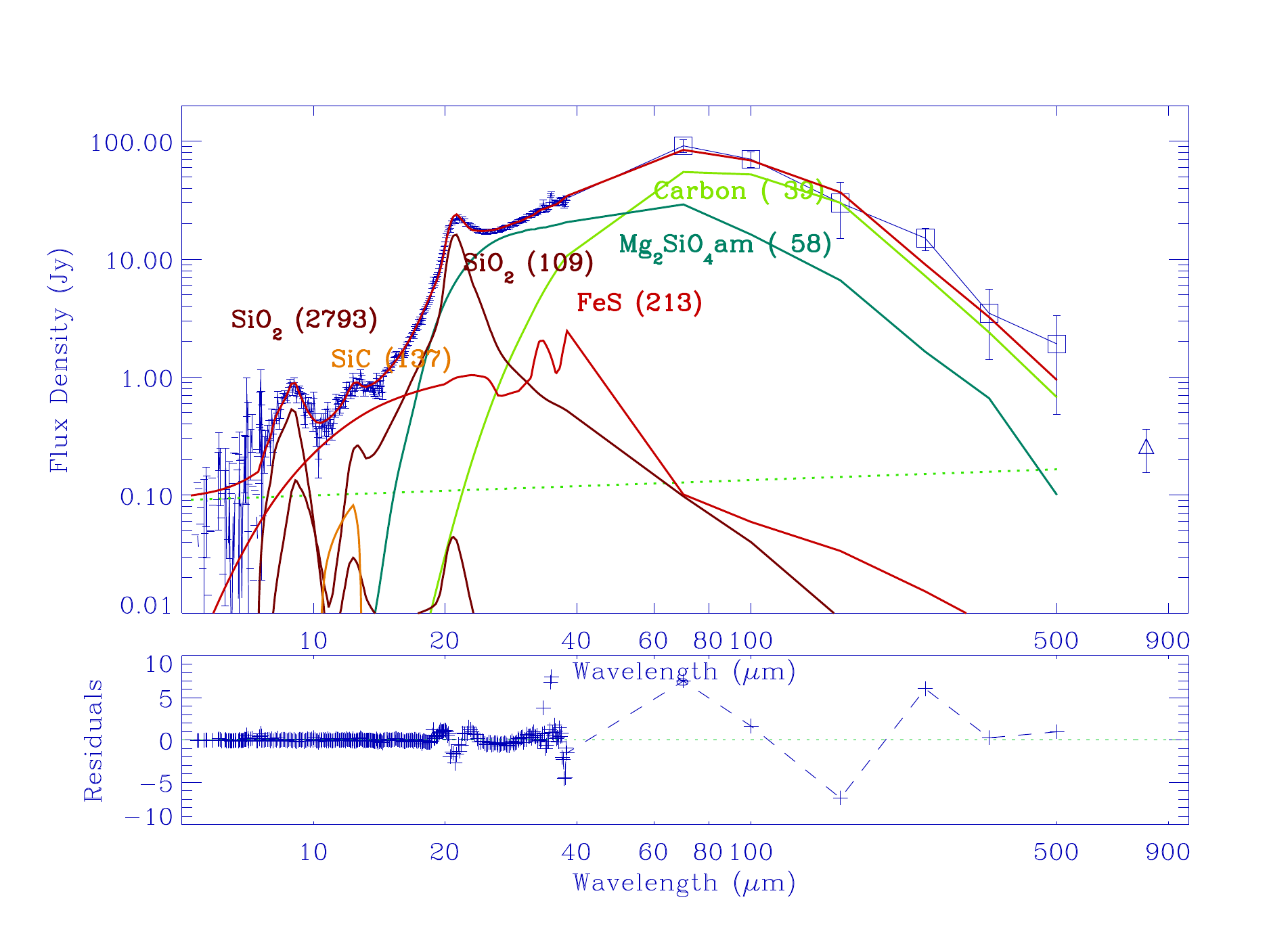}
\includegraphics[width=13truecm]{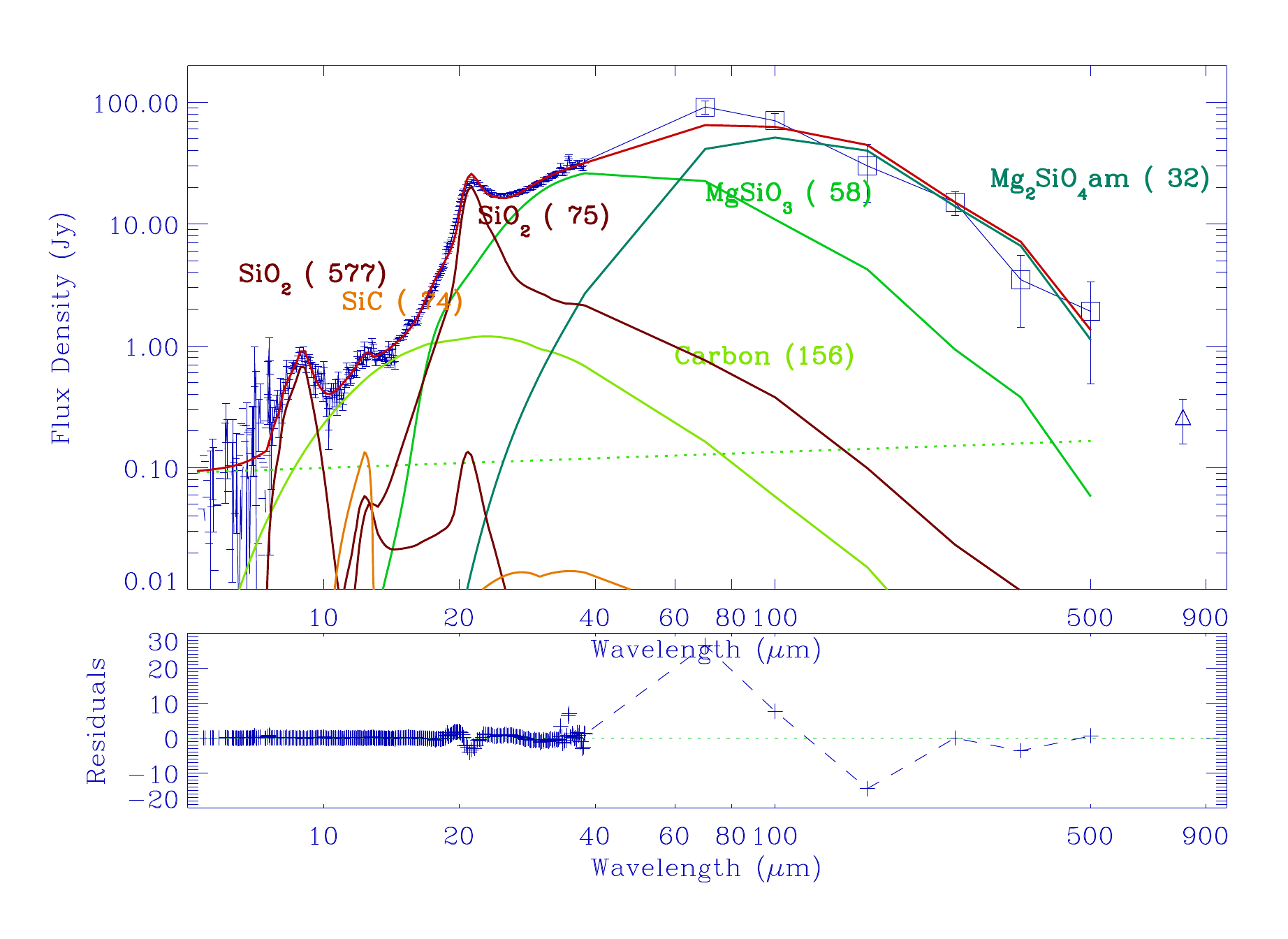}
\caption{Spectral energy distribution of G54.1+0.3 superposed on the dust fit of
{\it Top:} Model A4, assuming the cool dust component is made out of carbon
and this fit produces the best fit after accounting the errors as weights.
{\it Bottom:} The dust fit of Model B2; this is the same as Model A2 except
that carbon is assumed for the warm dust instead of FeS. The set of Model A
with the warm dust of FeS produces a better dust fit than those of Model B
with the warm dust of carbon.  }
   \label{fig:g54modelfitsb}
   \end{figure*}

Here we compare the effect of assuming different grain properties and grain
compositions. The results from all the models tested in this Section are
listed in Table~\ref{tab:Tdustmass}, along with the grain compositions
assumed, and the resultant dust temperatures and masses.  Some of the model
fits and their residuals are shown in Figures~\ref{fig:g54modelfitsa} and
\ref{fig:g54modelfitsb}. The 21$\mu$m and 11$\mu$m dust features are well
produced by the CDE grain model of silica and Silicon carbide (SiC),
respectively. As well as the dust features and compositions required to fit
the IRS spectrum of the SNR, we require a population of colder dust grains
to explain the fluxes at wavelengths $>$ 70 $\mu$m (Section
~\ref{sec:sedfitting}). 

In Model A1, the fit is produced with MgSiO$_3$ as a warm dust component,
Al$_2$O$_3$ as a cold dust componen and with SiO$_2$ to reproduce the
21$\mu$m feature. We fit the same Model A1 to the spectrum of Cas A. The
fit shown in Figure~\ref{fig:casafitModelA1} yielded a dust mass of 0.06
M$_\odot$, a 60\% higher dust mass than that from Model B (see Table 1) by
\cite{rho08}. Further work can explore the dust mass using SiO$_2$ and also
including the long-wavelength band Herschel data. In Model A2, we change
the cool dust composition in Model A1 from Al$_2$O$_3$ to Mg$_2$SiO$_4$ and
the warm dust from Mg$_2$SiO$_4$ to MgSiO$_3$ (see
Table~\ref{tab:Tdustmass}). The fit is poorer and generated a higher dust
mass ($0.9\,M_{\odot}$).   Next Model A3 uses MgSiO$_3$ (59\,K) and
Al$_2$O$_3$ (30\,K) instead of Mg$_2$SiO$_4$ (59\,K) and Al$_2$O$_3$
(40\,K) resulting in a dust mass of $0.35\,M_{\odot}$. Again this is a
poorer fit, with residuals at 40-100$\mu$m and 300-500$\mu$m.

Next we checked the effect of using temperature dependent dust absorption
properties. \citet{dupac03} observed large variations of the spectral index
($\beta \sim 0.8 - 2.4$) for dust with a wide range of temperatures (11 to
80 K), and suggested that there is an intrinsic dependence of the dust
spectral index on the temperature (T-$\beta$ anticorrelation). The reasons
can be the change in grain sizes, chemical composition of the grains in
different environments and due to quantum processes. \citet{paradis11} also
showed the flattening of the observed dust SED at far IR wavelengths using
DIRBE, Archeops and WMAP, this results in an increased emissivity.
Furthermore, \citet{demyk13, demyk17} and \citet{coupeaud11}  suggested
that the absorption coefficient of amorphous silicate grains decreases with
the temperature and shows a complex shape with the wavelength depending on
the micro structure of the materials. For wavelengths shorter than $\sim$
500 $\mu$m the spectral index is in the range 1.6-2.3 whereas at longer
wavelengths it changes and its value may be outside this range. To
determine if grains with temperature-dependent absorption properties could
fit the SED of G54.1+0.3, we used two datasets for grain compositions
closest to Mg$_2$SiO$_4$: one corresponds to glassy grains
Mg$_{1.9}$SiO$_{3.9}$ (``X35" sample) and the other corresponds to a more
porous sample of composition Mg$_{2.3}$SiO$_{4}$ (with a slight Mg excess,
``F1" sample). The spectra of these two analogues are different because the
analogues differs in terms of structure at micro- (nano-) meter scale, of
porosity, of chemical composition and homogeneity at small scale. Figure
\ref{fig:Qabsmg2sio4Tempe} shows the optical constants of these materials.
The Q$_{abs}$ depart from $\beta$=2 at $>$ 200$\mu$m as shown in Figure
\ref{fig:Qabsmg2sio4Tempe}. When Q$_{abs}$ decreases at low temperature,
the emissivity is flatter, and the flux is higher than expected with $\beta
=2$. Models A2F1 and A2X35 therefore include Mg$_{2.3}$SiO$_{4}$ and
Mg$_{1.9}$SiO$_{3.9}$ at a temperature of 30 K instead of Mg$_2$SiO$_4$.
Since Mg$_2$SiO$_4$ has lower Q$_{abs}$ at long wavelengths, this could
result in a higher dust mass for the SNR. The fits for these models (Table
\ref{tab:Tdustmass}, Figure~\ref{fig:g54modelfitsa}) were not as good as
the model with cold dust grains of Mg$_{2}$SiO$_{4}$ (Model A2) or
Al$_2$O$_3$ (Model A1).  We note that Q$_{abs}$ of Mg$_{2.3}$SiO$_{4}$ and
Mg$_{1.9}$SiO$_{3.9}$ show a factor of a few higher than that of
Mg$_{2}$SiO$_{4}$ above 40 $\mu$m; this likely resulted in the poorer fit
to the data.

Although theoretical models of dust formation in SNe \citep{nozawa03,
todini01} predicted that FeS grains are produced in the ejecta, we do not
detect the sharp 34$\mu$m dust feature in the spectra of G54.1+0.3 and Cas
A. We therefore try models where we replace FeS with carbon dust (with
featureless absorption coefficient).  First we replace the cold dust with
carbon in Model A4. This provides one of the best fits to the observed data
($\Delta \chi^2=3.7$) with a total dust mass of $0.26\,M_{\odot}$. We also
refit Models A1, A2 and A3 using carbon dust as the warm component (Models
B1, B2, and B3 - Table \ref{tab:Tdustmass}) but find that the quality of
the fits are slightly worse than those using FeS
(Figure~\ref{fig:g54modelfitsb}).

In order to reduce the relatively large residuals at medium or long
wavelengths, we also included additional warm or cold dust components using
MgSiO$_3$ or Al$_2$O$_3$, but the fits were not improved.
We have fitted with SiO$_2$ grains as a cool dust
component and the fit produced a large amount of dust of $sim$2.75 M$_{\odot}$
(see Table \ref{tab:Tdustmass}). However, this dust mass of SiO$_2$ is a
factor of two larger than the SiO$_2$ dust mass constraints predicted by
nuclear synthesis and dust formation models
\citep{nozawa03,cherchneff10,sukhbold16,temim17}; the predicted maximum
SiO$_2$ dust is about 1.3 M$_{\odot}$ for the largest possible progenitor
mass of 35 M$_{\odot}$. Thus, this model is a non-physical model and we
excluded this model from the total dust mass estimates. We attempted to fit
the spectra with two components of cool dust with the first component being
SiO$_2$, and the second component being Mg$_2$SiO$_4$, Al$_2$O$_3$, or
carbon dust, but the new fittings were not successful; the fits were either
statistically insignificant when we accounted for the errors, or the
reduced $\chi^2$ were too large. This is likely due to lack of far-IR data
points.

Table \ref{tab:Tdustmass} shows a summary of dust spectral fitting. The
individual fit includes 6 components: hot and warm silica components 
(with temperature ranges of 200-2700 K and 70-115 K), Silicon carbide,
two warm dust components (with temperature ranges of 150-210 K and
54-60 K), and a cool dust component (with a temperature range of 27-44 K). We
also included grains of SiO$_2$ at high temperature (300\,K) to account for
the emission feature around 9.0\mic, which was added to
mimic IR emission from stochastic heating (eg \citealt{lagage96, rho08}).

What did heat those dust grains$?$ For the case of Cas A, the dominant
heating source is a reverse shock. Since the dust is heated in young SNRs by the
reverse shock, it was easier to be detected than that in SNe where a
reverse shock has not yet passed SN ejecta (see Section \ref{sec:dustmass}
for details). When the gas temperature reaches about 1200 K  that was
$\sim$400 days after the SN explosion, dust forms and continues to cool
down for 3-5 years, the exact time scale and temperature depending on the
grain composition \citep[see][]{nozawa03}. The dust in SNe is too cool to
be detected or it is hard to distinguish the cool dust in SN ejecta from SN
light echoes.

\cite{koo08} showed a few OB stars in the field of
  the SNR G54.1+0.3. \cite{temim10, temim17} and \cite{kim13} suggested that
  they belong to the same Stellar Cluster as the progenitor of the SNR
  G54.1+0.3. Using near-IR spectroscopic observations, \cite{kim13}
  showed that the stars are O9 (17Msun) and a few early B stars, and
  suggested the progenitor of G54.1+0.3 to be Type IIP with a mass
  range of 8-20 M$_\odot$ ($<<$ 35M$_\odot$), so the progenitor is
  likely not a Wolf-Rayet star. Later O and early B type stars have
  temperatures of 37,000 K to 15,000 K. Herschel observations of
  such stars show a dust temperature of $\sim$ 60-80K, which
  contribute a small amount of the dust mass. Therefore, the OB stars
  may be the source of the warm dust temperature, but it is not clear
  if they are the heating source of the cold temperature component of
  the dust (20-40K) in the SNR. The analogy of G54.1+0.3 to the Crab
  Nebula discussed below may suggest that OB stars alone may not be
  unique heating sources of the dust in G54.1+0.3, although the
  possibility cannot be ruled out with current evidence.

For the Crab-like SNRs, the heating source is different because the
evidence of forward shock is unclear and thus the reverse shock may have
not developed \citep{hester08}. Since G54.1+0.3 is a Crab-like SNR, we
examined the heating mechanism of the Crab Nebula. The pulsar wind nebula
of the Crab Nebula  emits synchrotron radiation, where electrons are spinning
around the magnetic field. The existence of this shock driven by the
synchrotron nebula is recognized by Rayleigh-Taylor fingers
\citep{sankrit98} who estimate a shock velocity of 230 km s$^{-1}$, labeling the
shock as nonradiative. Numerical simulations of the expansion of the Crab Nebula
invariably identify these observed arcs with material from a surrounding
freely expanding remnant that has been compressed by the synchrotron-driven
shock \citep{jun98, bucciantini04}. The wind shock thermalizes flow energy,
accelerating electrons and positrons to energies as high as $\sim$10$^4$
TeV.

The dominant heating source for the dust in the Crab
Nebula is known to be synchrotron radiation (i.e. nonthernal radiation
field) in the pulsar wind nebula that depends on the luminosity of the
synchrotron radiation \citep[][]{davidson73, dwek81, dwek96, temim13}.
Grains are also heated by collisions between the electrons and the gas in
the filaments, but this heating is suggested to be insignificant in the
Crab-like SNRs \citep{hester08, dwek81,temim13, temim12}. We are suggesting the
heating source of G54.1+0.3 is similar to that of the Crab Nebula, dominant
heating being photoionization from synchrotron emission, synchrotron-driven
shock and possible contribution from collisional heating. Our results do
not support the claim that the stellar members of the SN progenitor's
cluster in G54.1+0.3 are the primary heating sources for the SN dust by
\cite{temim17} because we do not detect excess of the temperature and dust
mass at the position of stars (see Figure \ref{fig:g54dustmaps}). The blob
with the high temperature on the western shell (Figure
\ref{fig:g54dustmaps}a) coincides with the bright blob on the western shell
in the 24$\mu$m image and does not coincide with stars in the cluster. The heating
source of G54.1+0.3, which is responsible for the warm and cool grains we
observe, is likely photoionization from the synchrotron emission in the
G54.1+0.3's pulsar wind nebula.

\section{Discussion}
\subsection{Pre-solar Grains of Silica Produced in Supernovae}

Pre-solar grains are those formed before our Solar system in ancient
stellar outflows or ISM or supernovae and they are recognized by their
highly unusual isotopic compositions relative to all other materials (e.g.
Zinner 2013). Silicon carbide (SiC) is the best-studied type of presolar
grain and has been suggested to be formed in supernovae. Another well-known
pre-solar grain is corundum (Al$_2$O$_3$), low-density graphite (e.g.
carbon grains), amorphous silicates, forsterite and enstatite,  and
corundum \citep{messenger06}. Some isotopic anomalies of heavy elements,
especially $^{16}$O-enrichments, in meteorites have been attributed to the
dust that had condensed deep within expanding supernovae \citep{clayton04}.
The existence of isotopic anomalies in a major element within meteoritic
solids provided a decade of motivation for the search for supernova
stardust. Meteoritic studies have shown that some types of presolar grains
condense in the dense, warm stellar winds of evolved stars  but other types
of presolar grains (such as SiC-X grains) condenses in supernova explosions
\citep{clayton04}. Mantles on grains in molecular clouds are vaporized as
stars form in the clouds. A small fraction of dust survives planet
formation without alteration, protected inside asteroids.

We show that 10-13$\mu$m dust feature observed in G54.1+0.3 and Cas A are
attributed to Silicon carbide using two methods. We showed that spectral
fitting with the CDE model of SiC can reproduce the 11$\mu$m dust feature
again both for G54.1+0.3 and Cas A (see Figures~\ref{fig:G54fitModelA1},
\ref{fig:casafitModelA1}, \ref{fig:g54modelfitsa} and
\ref{fig:g54modelfitsb}). We suggest that SiC is responsible for the broad
dust bump between 10-13$\mu$m whereas the PAH emission is responsible for
the features at 11.0, 11.2 and 12.7\mic\ (see Section~\ref{sec:11peak}).

\begin{table*}
\caption[]{Summary of Dust Mass from Young Supernova Remnants.  $^a$ - The upper limit of warm dust mass estimated using 
\spitzer\ data only. Publications are: (1) Matsuura et al. (2015); (2) De Looze et al. (2017); 
(3) Rho et al. (2008); (4) Gomez et al. (2012);
(5) \cite{owen15}; (6) \cite{temim12}; this paper ($^b$see the text for detailed discussion and comparison with Temim et al. 2017); (7) Temim et al. (2010).}
\begin{center}
\begin{tabular}{llllll}
\\
\hline \hline
SNR & dust mass [\spitzer$^a$] & dust temperature & dominant & progenitor &ref. \\
    & (M$_{\odot}$) & K                    & dust type  & (M$_{\odot}$) \\
\hline
SN1987A & 0.4-0.7 [$>$1.3$\times$10$^{-3}$] &  17-23  & carbon & 18-20: Type II   & 1 \\
Cas A   & 0.3-0.5 [0.054]  & 30             & silicate & 25-30: Type II/Ib & 2, 3 \\
Crab Nebula & 0.1-0.54 [0.012] & 25-36      & carbon & 10-12: Type II  & 4, 5, 6 \\
G54.1+0.3 & {0.08-0.9} [0.02] & 27-44
         & silicate & 15-35: Type II &  this paper$^b$, 7 \\
\hline \hline
\end{tabular}
\end{center}
\label{tab:Tdustmassysnr}
\end{table*}

New presolar phases, such as SiO$_2$ have been identified with the Auger
spectroscopy to characterize the elemental compositions of pre-solar
silicate and oxide grains \citep{nguyen10}. They all exhibit enrichments in
$^{17}$O relative to solar, indicating origins in the envelopes of AGB
stars \citep{messenger06}. We showed that the 21$\mu$m dust feature in
G54.1+0.3 and Cas A can be reproduced with the CDE model of amorphous
silica (SiO$_2$). Theoretical models by \citet{nozawa03}  predicted silica
dust grains in supernova ejecta as well as other silicate dust such as
enstatite (MgSiO$_3$) and forsterite (Mg$_2$SiO$_4$) as our spectral fitting
also supports. Following \citet{schneider04}, \citet{bianchi07} have added
the formation of SiO$_2$ grains. We presented the first results of fitting
using SiO$_2$ to reproduce the 21$\mu$m dust feature in Cas A in conference
proceeding \citep{rho09proc}. Inspired from that, \citet{haenecour13}
searched SiO$_2$ from the primitive carbonaceous chondrites LaPaZ 031117
and Grove Mountains 021710 from meteorites, and they discovered two SiO$_2$
grains that are characterized by moderate enrichments in $^{18}$O relative
to solar and originated in Type II supernova ejecta. This proves that
SiO$_2$ is also pre-solar grains which are originated from supernovae. The
discovery of these two silica grains in laboratory provides support that
the silica dust which re-produces the 21$\mu$m dust feature in the young
SNR G54.1+0.3 and Cas A. \citet{arendt14} suggested that magnesium
silicates (particularly Mg$_{0.7}$SiO$_{2.7}$ have peaks that are in
roughly the right locations, but their shapes are not a good match to the
IRS data of Cas A, although two-composition fits help to improve the fits.
This particular silicates are not predicted by theoretical models of dust
in SN ejecta and our SiO$_2$ produces a better fit than that. Note that our
Q$_{abs}$ of SiO$_2$ is different from that in \citet{arendt14}.

The dust mass of silica depends on the progenitor mass, and
different theoretical models predict different amounts of dust mass. The
progenitor of G54.1+0.3 has a large range between 15-35 M$_{\odot}$ (see
Table \ref{tab:Tdustmassysnr}). The observed SiO$_2$ dust masses range
from 10$^{-5}$ to  2$\times$10$^{-3}$ M$_{\odot}$. 
Theoretical models \citep{nozawa03} predicted that for the unmixed
case the dust mass of SiO$_2$ is 0.02 - 0.2 M$_{\odot}$ ( 5\% - 12\% of the
total dust mass), and for 
mixed case the dust mass of SiO$_2$ is 0.6 - 1.3 M$_{\odot}$ (70\% - 50\%
of the total mass). The other dominant dust mass is predicted to be
from Mg$_2$SiO$_4$. \cite{cherchneff10} predicted that the dust mass of
SiO$_2$ is 67\% (about 0.1 M$_{\odot}$) of the total dust mass for the no
depletion case, while very small percent of SiO$_2$ is formed for the Mg/Fe
depleted case. \cite{sarangi13} suggest a small amount of SiO$_2$, one order
of 10$^{-5}$ M$_{\odot}$ as we observed for most of our spectral models. In order to
accurately characterize a dozen dust species that are predicted by the
theoretical models, a higher number of far-IR data points or far-IR
spectroscopic data are required. Theoretical models of small grids in the
progenitor mass and other physical parameters that determine the dust mass
will also be helpful.

\subsection{The dust mass in supernova ejecta and its implication to the dust in the early Universe}
\label{sec:dustmass}

The dust mass we estimated from G54.1+0.3 is {\bf 0.08-0.90$\,M_{\odot}$}
(see Table \ref{tab:Tdustmass}) depending on the grain composition, which
is comparable to predicted masses from theoretical models. The dust mass
estimate of G54.1+0.3 can be more accurate than that of Cas A because of
the following two reasons; (1) the foreground ISM contribution is less
obvious than that of Cas A \citep{barlow10} in far-IR emission as shown in
Figure \ref{fig:g54herschelimagecolor}, and (2) the dust destruction due to
the forward or reverse shock is less significant because the SNR  G54.1+0.3
is a Crab-like SNR. 
G54.1+0.3 and Cas A have typical foreground emission of 0.08  and 0.32 Jy
(9$''$pixel)$^{-1}$ at 160$\mu$m, and 0.3 and 0.55 Jy beam$^{-1}$ at
250$\mu$m, respectively. The surface brightness of the foreground emission
in vicinity of G54.1+0.3  is $\sim$20-30 percent of the SNR brightness,
whereas the typical brightness of foreground emission near Cas A
is $\sim$90 - 115 percent of the Cas A brightness. Most importantly, the filamentary
cirrus structures continue to be present across the SNR Cas A in the {\it
Herschel} images \citep[see Figure 1 of][]{barlow10}, while such
cirrus structures are not obvious toward G54.1+0.3 (see Figures
\ref{fig:g54herschelimagecolor} and \ref{fig:g54herschelimagenew}).
The statistical uncertainties of dust temperatures and
masses for each spectral fitting results are given  in Table
\ref{tab:Tdustmass}. The uncertainty of dust mass ranges between 7-35
percent.

Our estimated dust mass is somewhat lower (0.07-1.2 M$_\odot$ after
accounting for the errors) than that (1.1$\pm$0.8 M$_\odot$) obtained by
\citet[][]{temim17}. Considering completely independent methods, this is a
reasonable agreement since the systematic errors could be this large caused
by different grain compositions, absorption coefficiencies used and the
flux measurements of \herschel\ due to large background variations.

In Table \ref{tab:Tdustmass}, the dust mass of {\bf 0.08-0.90$\,M_{\odot}$}
primarily comes from the cool dust component ($>$90\%), and the warm dust
contributes to the dust mass less than 10\% except the case of of
Fe$_3$O$_4$ as a cool dust (model A5 and B4). Another noticeable conclusion
is when we use Mg$_2$SiO$_4$ as a cool component, the fitting (Models A1
and B2) gives higher dust masses than the other models. \citet{bianchi07}
show that large amounts (0.1 M$_\odot$) of cool dust (T $\sim$ 40 K) can be
present which is similar to what we observed within a factor of 3, and 
for the case of olivine Mg$_2$SiO$_4$ as a cool dust component 
we observed a large dust mass (up to 0.88 M$_{\odot}$).

\begin{figure}
\includegraphics[width=8truecm]{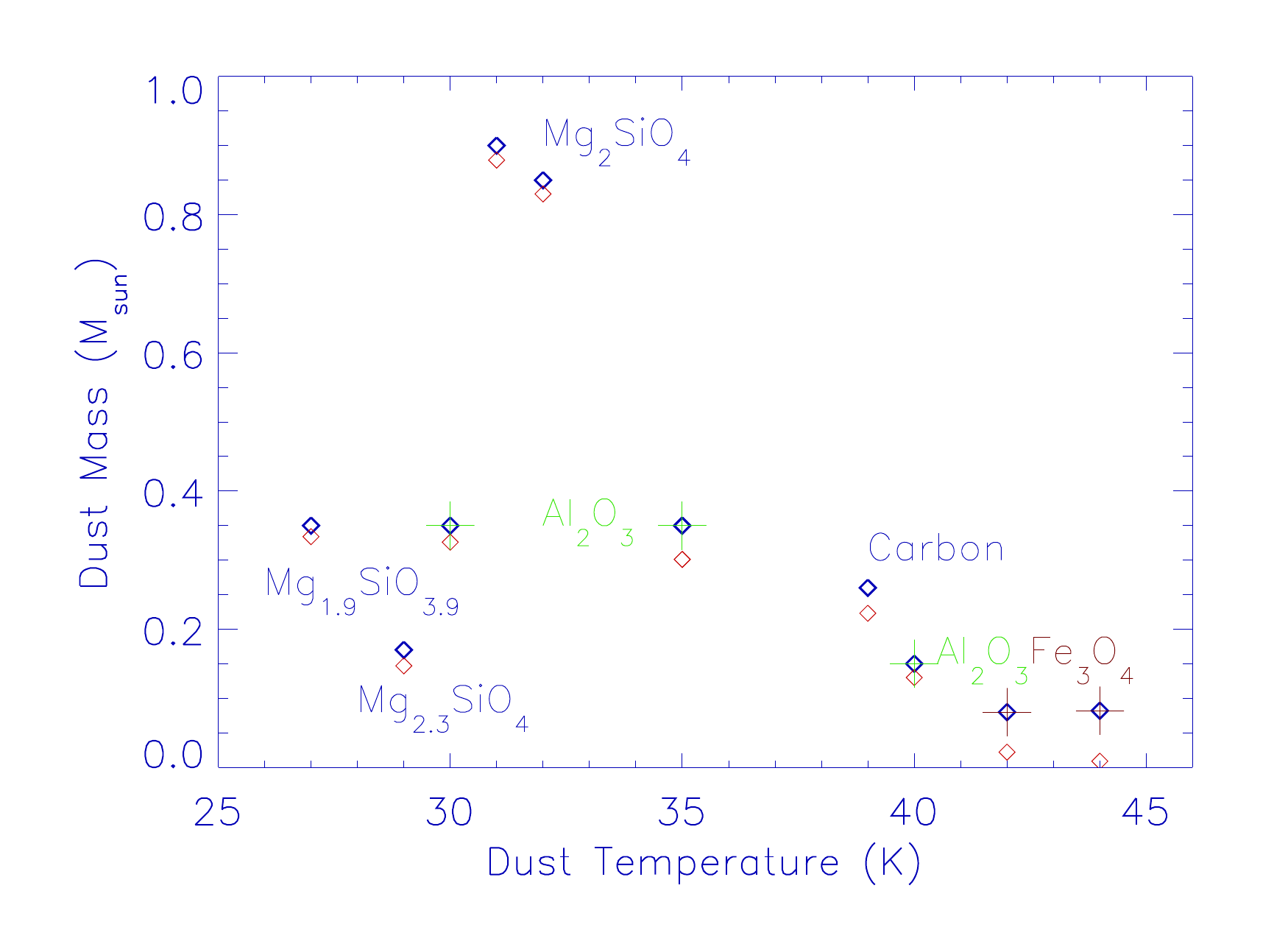}
\caption{The derived dust mass for dust grains at different dust
temperatures and assumed grain compositions. The total dust mass and the
dust mass of the cool component are marked as diamonds in blue and red,
respectively.}
\label{fig:dusttempmass}
\end{figure}

We examined the correlation between the derived dust mass and the fitted
dust temperature of the cool component. There is an expectation that dust
mass is higher when the dust temperature is lower  as the cool dust dominates its dust
mass as shown in Table \ref{tab:Tdustmass}. \citet{gall11} also shows
this correlation in young SNe. Figure~\ref{fig:dusttempmass} shows
the distribution of dust temperature versus mass for the cool dust
component for G54.1+0.3 given the compositions assumed in this work. 
The dust mass is inversely proportional to the Planck function which is a
function of temperature as shown in the equation in Section
\ref{sec:sedfitting}. The trend is observed in
Figure~\ref{fig:dusttempmass} when we exclude Mg$_2$SiO$_4$. The dust mass
depends on not only the temperature but also the density and absorption
coefficient of the grain composition. The cold grains dominate the dust
mass as shown in Table \ref{tab:Tdustmass}. However, for some cases, the
dust mass of warm dust was higher than that of cool dust (see the case of
Model A5) where the warm dust reproduces the large portion of continuum and
has low Q$_{abs}$ compared with those of other grains (see Fig.
\ref{fig:Qabsmg2sio4Tempe}).
The total mass of dust we derive here for a range
of grain compositions is between 0.08-0.9\,M$_\odot$ including
temperature-dependent silicate grains.  The one exception is when we assume
grains composed of forsterite (Mg$_2$SiO$_4$) which yields a higher dust
mass than other models. Our results show that the grain composition is also
an important factor as well as the temperature to determine the total dust
mass (as well as the cool dust temperature).

Whether some of the dust formed in the ejecta will survive is a major
unresolved issue.  The lifetimes of the major dust materials (i.e. carbon
and silicates) against the rate of destruction by SN shock waves appear to
be too short compared to the time-scales for the formation of new dust
\citep{jones94}. Recent simulations by \citet{slavin15} shows that the dust
destruction timescales are increased by a factor of 2-6 times compared to
those of \citet{jones96} when accounting for the thermal history of a
shocked gas and longer silicate grain destruction timescale. \citet{nath08}
find that a fraction of the dust mass (1\%- 20\% for silicates and
graphites) can be sputtered by reverse shocks, the fraction varying with
the size of ejecta knots, the grain size distribution and the steepness of
the density profile of the ejecta mass. \citet{silvia10} find that for high
ejecta densities, the percentage of dust that survives is strongly
dependent on the relative velocity between the clump and the reverse shock,
causing up to 50\% more destruction for the highest velocity shock. The
rates of dust destruction therefore depends on the pre-shock density,
progenitor mass, shock speed and so on. At least 2-20\% of dust is survived
behind the reverse shock \citep{bianchi07, micelotta16} where knots can be
as dense as 10$^{8}$ cm$^{-3}$ (based on CO observations for Cas A
\citealp{rho09co,rho12}). For G54.1+0.3, dust destruction by a forward shock
may be less critical, since the evidence of strong forward shock has not
been observed in the Crab-like SNRs \citep{hester08}.

Our observations of G54.1+0.3 implies that {\bf 0.08-0.9 M$_{\odot}$ of
dust is formed in the ejecta} with cool temperatures ranging from 27-44\,K.
This  mass is consistent with the predicted dust mass from the theoretical
models of dust formation in SN ejecta \citep{nozawa03, todini01,
bianchi07}. The estimated dust mass from the G54.1+0.3 implies that SNe are
a significant source of dust in the early Universe. Table
\ref{tab:Tdustmassysnr} compares the dust masses observed with \spitzer\
alone and a combination of \spitzer\ and \herschel\ observations. The dust
masses observed with Herschel are an order of magnitude larger than those
previously measured in these remnants using {\it Spitzer}, and are two
orders of magnitude higher than those observed in young SNe
\citep{kotak06,gall11}.  As with SN1987A, Cas A and the Crab Nebula (see
Table~\ref{tab:Tdustmassysnr}), we find that the mass of dust in the
G54.1+0.3 SNR is very different based on SED fitting with and without
adding the FIR-submm regime. Spectral fitting including the long wavelength
SED in Figure~\ref{fig:G54fitModelA1} requires a significant amount of dust
at lower temperatures.  Thus, this work adds to the growing evidence that
core-collapse SNe are forming large quantities of dust.

Future JWST observations of SNe will help to resolve if the dust observed
in SNe is from ejecta, dense shell from stellar wind, or from light echos,
and will allow us to observe dust formation of different types of dust from
300 - 800 days after the explosion.

\section{Conclusion}

\begin{enumerate}
 \item We report the detection of far-infrared and submillimeter emission
 from the Crab-like SNR G54.1+0.3 using SHARC-II, LABOCA and \herschel\
 observations (70-500\mic).

 \item Using \spitzer\ archival IRS, IRAC, and MIPS data, we detected a 21\mic\
dust feature in G54.1+0.3 remarkably similar to that seen in Cas A. As in
Cas A, this 21\mic\ dust feature is coincident with the ejecta line
emission as traced by [Ar II]. SED fitting suggests that the 21$\mu$m
feature is composed of SiO$_2$ grains. The 21$\mu$m feature in the SNRs
G54.1+0.3 and Cas A is offset by $\sim$1$\mu$m from the dust feature
(peaking at 20.3$\mu$m) in the PPNe or carbon stars, and neither SNR has a
broad dust bump at 30$\mu$m which is seen in the PPNe. The 21$\mu$m dust
feature, therefore, seems to be unique to Cas A and G54.1+0.3 to date.

 \item We suggest that the 21$\mu$m dust features in G54.1+0.3 and Cas A are
attributed to the presolar grains of SiO$_2$. We raise a possibility that
the 11$\mu$m dust features may come from the pre-solar grain of SiC with a
combination of PAH contribution.

\item We reveal a cold dust component (27-44\,K) coincident with the SNR
with a total dust mass of {\bf $0.08-0.9\,M_{\odot}$} depending on the
grain composition within the ejecta.  If the FIR and submm emission is
indeed dust formed in the ejecta, this makes G54.1+0.3 one of only four
SNRs where cold dust is detected. The amount of dust mass observed from the
SNR G54.1+0.3 suggests that SNe are a significant source of dust
in the early Universe.

\end{enumerate}

We thank anonymous referee for insightful comments, which helped to improve the paper.   
JR acknowledges support from NASA ADAP grant (NNX12AG97G which was rolled
over from NNX10AQ84G) for the study of SN dust. HLG acknowledges support
from the European Research Council (ERC) in the form of Consolidator Grant
{\sc CosmicDust} (ERC-2014-CoG-647939, PI H\,L\,Gomez). We thanks Karine
Demyk for providing absorption coefficients prior to her publication,
Mikako Matsuura for information and discussion about SN1987A and dust in
SNRs, and William Reach for helping calculation of mineral properties. JR
thanks Achim Tappe to reduce the IRS spectra for proto-planetary nebulae. 




   \bibliographystyle{mnras}
   \bibliography{msrefs} 






   \bsp	
   \label{lastpage}
   \end{document}